\documentclass[%
aip,
jcp,
amsmath,amssymb,
preprint%
]{revtex4-2}
\usepackage{amsmath,amssymb,amsthm, mathtools, enumitem, empheq, gensymb, textcomp, bm, booktabs, multirow}
\usepackage[margin=1in]{geometry}
\usepackage{chemformula}
\usepackage{dsfont}{\tiny }
\usepackage{algorithm}
\usepackage[noend]{algpseudocode}
\usepackage[normalem]{ulem}
\usepackage{float}

\def\bea{\begin{eqnarray}}
	\def\eea{\end{eqnarray}}
\def\nn{\nonumber}
\def\beq{\begin{equation}}
	\def\eeq{\end{equation}}
\def\ba{\begin{eqnarray}}
	\def\ea{\end{eqnarray}}
\def\be{\ba\displaystyle}
\def\ee{\ea}

\definecolor{cadmiumgreen}{rgb}{0.0, 0.42, 0.24}
\definecolor{carmine}{rgb}{0.59, 0.0, 0.09}

\newcommand{\la}{\langle}
\newcommand{\ra}{\rangle}

\DeclarePairedDelimiter\ket{\lvert}{\rangle}
\DeclarePairedDelimiterX\braket[2]{\langle}{\rangle}{#1 \delimsize\vert #2}
\DeclarePairedDelimiterX\Braket[3]{\langle}{\rangle}{#1 \delimsize\vert #2 \delimsize\vert #3}

\begin{document}
	
	
	\title{Advantage of discrete variable representation in variational quantum eigensolvers for vibrational energy calculations}
	\author{K. Asnaashari}
	\email{kasra.asnaashari@phys.chem.ethz.ch}
	\author{D. Bondarenko}
	\author{R. V. Krems}
	\affiliation{ 
		Department of Chemistry, University of British Columbia, \\Vancouver, B.C. V6T 1Z1, Canada \\
		Stewart Blusson Quantum Matter Institute, \\
		Vancouver, B.C. V6T 1Z4, Canada
	}
	
	\date{\today}

	\begin{abstract}
		While quantum computing algorithms have been widely applied for electronic structure calculations, applications to molecular dynamics remain scarce. 
		Complex and varied landscapes of molecular potential energy surfaces give rise to vibrational states with a wide range of properties, making it difficult to construct a general representation of ro-vibrational states by a quantum computer with a limited number of qubits and gates. Another challenge is the exponential growth of the computation complexity -- for example, the number of terms required to expand a general Hamiltonian in Pauli strings increases exponentially with the number of qubits. 
		Here, we show that discrete variable representation (DVR) can be leveraged to represent molecular Hamiltonians by the polynomial (in the number of qubits) number of quantum circuits. 
		We then demonstrate that DVR Hamiltonians lead to very efficient quantum ansatze for vibrational states. For this purpose, we develop a compositional
		quantum ansatz search that adapts gate sequences in variational quantum eigensolvers (VQE) to a specific molecular state. We apply VQE to compute the vibrational energy levels of Cr$_2$ in seven electronic states as well as of van der Waals complexes Ar--HCl and Mg--NH. Our numerical results show that accuracy of 1~cm$^{-1}$ can be achieved by very shallow quantum circuits with 2 to 9 entangling gates.

	\end{abstract}

	\maketitle

	\section{Introduction}
	
	Accurate calculation of molecular properties is considered a promising application of quantum computing. 
	The eigenstates of molecular Hamiltonians can be obtained on quantum computers by variational quantum eigensolvers (VQEs) \cite{vqe-review, vqd} that employ sequences of gates operating on qubits (quantum circuits) to prepare quantum states tailored for specific problems. 
	VQEs have been applied for solving the electronic structure problem for molecules \cite{vqe-1,vqe-2,chem-1,chem-2,chem-3,chem-4,chem-5,chem-6,chem-7} and lattice models \cite{chem-3,vqe-review,lattice-1}. 
	However, applications of VQE to computations of ro-vibrational energies and states have been limited \cite{dvr-dynamics, morse, mol-vib, mol-vib-2, mol-vib-3, co2-1, co2-2, co2-3}. 
	Refs. \cite{mol-vib-3, mol-vib-2} demonstrated a general approach for computing ro-vibrational energy levels of polyatomic molecules inspired by previous work  on electronic structure. However, the methods of  Refs. \cite{mol-vib-2, mol-vib-3} require extended quantum circuits including a large number of entangling gates. This makes the applications of VQE to vibrational energy calculations challenging for quantum computers limited by errors and noise. The errors grow with the circuit size and make large quantum circuits impractical for implementation on current quantum devices.  This challenge is compounded by transpilation of quantum circuits into hardware-specific gate sequences, which often further extends the size  of the quantum circuits.

	A central goal for quantum computing of molecular dynamics can thus be formulated as to develop a general approach that (i) is applicable to a broad range of molecules with widely varying ro-vibrational states, from deeply bound to van der Waals states; (ii) yields high accuracy with  shallow quantum circuits; (iii) exhibits at most a polynomial scaling with the dimensionality of the molecular configuration space or, alternatively, in the number of qubits if binary basis encoding is used. Within the VQE framework, requirement (i) demands an ansatz for quantum circuits that can represent molecular states in widely varying landscapes of potential energy surfaces; (ii) is necessary due to noise and hardware limitations of current quantum computers; and (iii) is essential for realizing the quantum advantage over classical methods. 
	
	
	Here, we explore an approach to computing the vibrational energy levels of molecules with VQE based on discrete variable representations (DVR) of molecular Hamiltonians \cite{colbert, choi}. 
	We first derive the theoretical bounds on the quantum measurement complexity with VQE based on Fourier grid DVR.  
	Our analysis shows that 
	the structure of the DVR matrices can be exploited to reduce the expansion of the molecular Hamiltonians in measurement operators 
	to the desired polynomial scaling in the number of qubits.
	In the second part of this work, we explore an automated construction of the quantum circuits for the vibrational energy computations. 
	Our goal is to build quantum circuits, yielding accurate VQE results, without any constraints on the quantum ansatz. 
	Our results demonstrate that DVR Hamiltonians lead to very efficient (small number of qubits and gates) quantum circuits for representing vibrational states of molecules by states of a quantum computer. 
	To probe the minimal gate count for accurate vibrational VQE computations, we develop a compositional search algorithm that incrementally grows quantum circuits to identify optimal sequences in the space of gate permutations.

	To illustrate the generality of this approach and the efficiency of the resulting quantum circuit representations of vibrational states, we consider Cr$_2$ 
	in seven different electronic states \cite{cr2-pot} and van der Waals complexes Ar--HCl($^1\Sigma$) and Mg--NH($^3\Sigma$). These molecular systems exhibit vibrational states with widely different energies (from $-55$ to $-15,000$ cm$^{-1}$ from the dissociation threshold) and spatial variations of wave functions and energy level patterns. 
	Our compositional search yields quantum circuits that produce VQE results with accuracy $<1$ cm$^{-1}$ for ground and excited vibrational energy levels, illustrating the ability of VQE to compute the rotational constants and vibrational anharmonicity, with between 2 and 9 entangling gates, for diatomic and triatomic molecules.   	
	For reference, previous VQE calculations of vibrational energy levels required extended quantum circuits with $>200$ (for CO, COH and O$_3$ molecules  \cite{mol-vib-3}) or between 44 and 140 292 (for CO$_2$, H$_2$CO and HCOOH molecules \cite{mol-vib-2}) entangling gates.

	The remainder of this article is organized as follows. After a brief introduction of VQE, Section \ref{efficient-measurements} presents an algorithm to evaluate the DVR Hamiltonains with a polynomial number of measurements. 
	Section \ref{anzats-optimization} describes the algorithm for the compositional anzats optimization, which is followed by numerical results illustrating the efficiency and accuracy of the optimized ansatze for Cr$_2$ 
	in seven different electronic states \cite{cr2-pot} and van der Waals complexes Ar--HCl($^1\Sigma$) and Mg--NH($^3\Sigma$) in Section \ref{numerical-results}. 
	The work is summarized in Section \ref{summary}. 
	
	\clearpage
	\newpage
	
	\section{Theory}

	In VQE, 
	a quantum computer estimates the expectation value  $\Braket{\psi(\bm\varphi)}{\hat{H}}{\psi(\bm\varphi)}$, which  is minimized by varying $\bm\varphi$ to yield 
	the lowest eigenvalue $E_{i=0}$ and an approximate representation of the corresponding eigenvector of $\hat H$. 
	This method can be extended to compute excited states  by optimizing \cite{vqd}
	\begin{align}
		\tilde{\bm \varphi}_v &= {\rm argmin}_{\bm\varphi} \left [ \Braket{\psi(\bm\varphi)}{\hat H}{\psi(\bm\varphi)} 
		+ \sum_{i=0}^{v-1} \beta_i \langle {\psi(\tilde{\bm\varphi}_i)}| {\psi({\bm\varphi})\rangle} \right ]
		\label{Hexcited}
	\end{align}
	where $\beta_i \geq E_{i+1} - E_i$ and $\tilde {\bm \varphi}_i$ denotes an optimal solution for the corresponding quantum state.

	The quantum states $| \psi({\bm \varphi}) \rangle$ are obtained by quantum circuits acting on qubits and the Hamiltonian is expanded in quantum operators. Most generally, 
	\begin{align}
		\hat H = \sum_{i = 0}^{4^n} A_i K^i_1\otimes K^i_2 \ldots \otimes K^i_n 
		\label{Pauli}
	\end{align}
	where $n$ is the number of qubits, $K^i_j \in \{\sigma_X, \sigma_Y, \sigma_Z, I\}$ acting on qubit $j$, $\{ \sigma_i \}$ are the Pauli matrices, $I$ is the identity matrix, 
	\begin{align}
		A_i = \frac{1}{2^n}{\rm Tr}[(K^i_1\otimes K^i_2 \ldots \otimes K^i_n)\cdot {\bm H}]
	\end{align}
	and $\bm H$ is the Hamiltonian matrix in some basis. 
	The computational complexity is determined by the number of non-zero terms in Eq. (\ref{Pauli}), which for a general matrix is $4^n$; 
	and the complexity of the quantum circuits yielding $\ket{\psi(\bm\varphi)}$.
	In section \ref{efficient-measurements}, we show that the structure of DVR matrices allows an efficient quantum circuit representation of $\hat H$,
	scaling with $n$ as poly($n$).

	\begin{figure}[h]
		\centering
		\includegraphics[width=0.5\columnwidth]{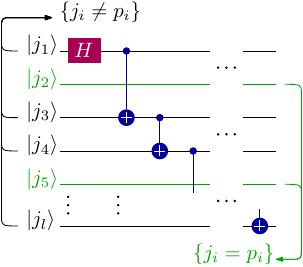}~~~~~\includegraphics[width=0.4\columnwidth]{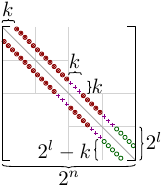}
		\caption{Left: Preparation of state (\ref{plus-l}) from the computational basis. $H$ denotes the Hadamard gate and the circles -- the CNOT gates. Right: Measuring $t^{k[n]}$ that includes matrix elements on the main and $k^{\text{th}}$ diagonals. The open circles $\color{cadmiumgreen} \circ$ are obtained from projections of (\ref{plus-l}),  $\color{carmine} \otimes $ -- via $\mathds{1}_2 \otimes t^{k[m]}$, and $\color{magenta}+$ -- by additional measurements in the entangled basis. }
		\label{fig:BellCirc}
	\end{figure}

	\subsection{Efficient measurement of DVR Hamiltonians}
	\label{efficient-measurements}
	DVR is a finite basis representation, in which the coordinate operators (and consequently the potential energy) are diagonal. 
	The DVR matrix of kinetic energy is not sparse. However, it has specific structure that is exploited here. 
	We use DVR introduced by Colbert and Miller \cite{colbert}.  Other DVR bases can be reduced to those in Ref. \cite{colbert} by coordinate transformations. 
	%
	As follows from Ref. \cite{colbert}, the Hamiltonian matrix in the DVR representation has the following structure: 
	\be
	H_{ij} =
	\begin{cases}
		d(i), &  i=j,\\
		f\left(|i-j|\right) + g(i+j), & i \neq j
	\end{cases}.
	\label{dvr-structure}
	\ee	
	We use a number encoding to map the DVR basis states onto qubit states, where the indices of the flattened DVR basis states are mapped into their binary representation on the qubits. This allows VQE to compute the eigenvalues of $\bm H$ of size
	$2^n\times 2^n$ using $n$ qubits. As the number of DVR bases required for accurate results grows exponentially with the number of degrees of freedom of the system, the number of qubits necessary to represent system grows polynomially with the size of the system.

	As shown in Appendix A, the functions $f$ and $g$ in  Eq. (\ref{dvr-structure}) satisfy
	\be
	\label{off-diagonals1}
	\sum_{k=s}^{2^n-1}\left| f(k) \right| \leq  O(s^{-\alpha}), \ \alpha>0,\\
	\sum_{k=r}^{2^{n+1}-1-r}\left| g(k) \right| \leq O(r^{-\beta}), \ \beta>0,
	\label{off-diagonals2}
	\ee
	for $1 \leq r \ll N = 2^n$ and $1 \leq s \ll N = 2^n$, where $N$ is the number of DVR states, assumed to be large. 

	Given a quantum state $|\psi \rangle$, our aim is to show that
	\be
	\tau = \la \psi| \hat H | \psi \ra + O(\epsilon) 
	\label{ev}
	\ee
	can be measured with the number of quantum circuits that scales polynomially with $n$ and $1/\epsilon$. 
	We first use Eqs. (\ref{off-diagonals1}) and (\ref{off-diagonals2}) to show that the expectation value (\ref{ev}) can be approximated as
	\be
	\tau = \la \psi| \hat H^{(s,r)} | \psi \ra + O(\epsilon) 
	\label{truncated-ev}
	\ee
	where $ \hat H^{(s,r)}$ employs a truncated DVR matrix with the following elements:
	\be
	H^{(s,r)}_{ij} = 
	\begin{cases}
		d(i) & \ i=j,\\
		f\left(|i-j|\right) + g(i+j) & \ i \neq j,\ |i-j|< s,\ i+j < r,\\
		0 \ \text{otherwise},
	\end{cases}
	\ee
	for 
	\be
	s \sim \epsilon^{-1/\alpha}
	\label{s}
	\ee
	and 
	\be
	r \sim \epsilon^{-1/\beta}.
	\label{r}
	\ee

	To prove Eq. (\ref{truncated-ev}),  we consider operators defined as a $k^\text{th}$ (anti)-diagonal matrix, with elements 
	\be
	K^{k \pm \pm}_{ij} \equiv \delta_{i \pm j, \pm k},
	\ee
	where only $++$, $+-$ and $--$ combinations of signs are used. Since $K^{k \pm \pm} $ either permute or shift vector components, while either preserving the norm of vectors or truncating vectors in a finite Hilbert space, 
	$\| K^{k \pm \pm} \psi \|_2 \leq \| \psi \|_2$ (and the inequality is tight). This can be used in the Cauchy–Schwarz inequality $|\la \psi, K^{k \pm \pm} \psi \ra| \leq \| \psi \|_2 \| K^{k \pm \pm} \psi \|_2$ to bound  the expectation value of $K^{k \pm \pm}$ as
	\be
	|\la \psi, K^{k \pm \pm} \psi \ra| \leq \| \psi \|^2_2.
	\ee
	Using the triangle inequality and $\| \psi \|^2_2=1$, the approximation error can be bounded by
	\be
	\left|\la\psi|\left(\hat H - \hat H^{(s,r)}\right) |\psi\ra \right| &=& 
	\left|\la\psi|\left(\sum_{k=s}^{2^n-1}f(k)\left[K^{k-+} + K^{k--}\right] + \sum_{k=r}^{2^{n+1}-1-r}g(k) K^{k++}\right) |\psi\ra \right|
	\nn\\ &\leq&
	\left( 2 \sum_{k=s}^{2^n-1}\left| f(k) \right| + \sum_{k=r}^{2^{n+1}-1-r}\left| g(k) \right| \right)
	\ee
	Due to Eqs. (\ref{off-diagonals1}) and (\ref{off-diagonals2}), Eq. (\ref{truncated-ev}) holds for  $s$ and $r$ given by Eqs. (\ref{s}) and (\ref{r}).

	We next seek to transform $|\psi \rangle$ by short-depth $\hat V_i$, 
	so that
	\be
	\tau = \sum_{i=1}^{{\rm poly}(n, 1/\epsilon)} \sum_{j=1}^{2^n} w_{ij} \left|\la \psi |V_i^{\dagger} | j \ra\right|^2
	\ee
	where $j$ is the index that enumerates the states of $n$ qubits in the computational $Z$ basis.  
	We leverage Eq. (\ref{truncated-ev}) to decompose $\bm \tau$ into contributions from a diagonal matrix ($D$), $s \approx \epsilon^{-1/\alpha}$ diagonal bands ($t^k$) and $r\approx \epsilon^{-1/\beta}$ anti-diagonal components ($a^k$), as follows: 
	\be 
	\label{the-tau}
	\tau = \langle D \rangle +  \sum_{k=1}^{\left\lceil \epsilon^{- \frac{1}{\alpha}} \right\rceil}  f(k) \la t^{k[n]} \ra +  ~~~~~~~~~~~~~~   \\  \sum_{k=1}^{\left\lceil \epsilon^{- \frac{1}{\beta}} \right\rceil}\left( g(k) \la a^{k[n]} \ra + g(2^n-k) \la a^{(2^n-k)[n]} \ra \right)  + O\left(\epsilon\right).
	\nonumber
	\ee
	We note that the construction of $t^{k[n]}$ and $a^{k[n]}$ described below produces contributions to the main diagonal upon measurements. This is offset by a corresponding change of the diagonal matrix elements, as specified in Appendix B. The computation of the diagonal contribution $\langle D \rangle$ is classically efficient. In this case, one can set $\{ V_i \} = \{ \mathds{1} \}$ and combine $w_{1i} = D_{ii}$ from Appendix B with the weights given by the measurements in the computational basis.

	
	To construct $t^{k[n]}$, we note that $ l \equiv \left\lceil \log_2(k+1) \right\rceil$ is the smallest number of qubits that allows $k$-th band.
	We consider $l$-qubit entangled states produced from the $Z$ basis by a sequence of CNOT gates shown in Fig. \ref{fig:BellCirc} (left). 
	The resulting state is
	\begin{eqnarray}
		| { {\bm j}, {\bm p } \ra}_l =  \frac{ | { {\bm j} \ra}  + | { {\bm p} \ra}  }{\sqrt{2}},
		\label{plus-l}
	\end{eqnarray}
	where $| { {\bm j} \ra} = \bigotimes_i^l |j_i\ra$ and $| { {\bm p} \ra} = \bigotimes_i^l |p_i\ra$ are the binary representations of the row ($p$) and column ($j$) indices of the DVR matrix, 
	with $p = j + k$,  and $j_i$ and $p_i$ representing single-qubit states.
	The projection of an $n$-qubit state onto the $| { {\bm j}, {\bm p } \ra}_l $ state 
	has 2 non-zero off-diagonal matrix elements of magnitude $1/2$. The binary representation of the row and column positions of these off-diagonal elements differ by $l$ bits. 
	The position of the non-zero off-diagonal elements is controlled by the position of the CNOT gates in Fig. \ref{fig:BellCirc}.  
	Thus, the expectation value of $t^{k[l]}$ can be obtained by measurements with, at most, $k$ $l$-qubit entangled states.


	We now observe that $t^{k[m+1
		]} = \mathds{1}_2 \otimes t^{k[m]} + \gamma_{k[m]}$, where $\gamma_{k[m]}$ represents $k$ pairs of elements missing from the middle of the tensor product matrix (shown as pluses in Fig.~\ref{fig:BellCirc} right). 
	These elements can be directly targeted by additional measurements in an entangled basis, analogous to states in Eq. (\ref{plus-l}) but with a different number $q$ of qubits, 
	\begin{eqnarray}
		| { {\bm j}, {\bm p } \ra}_{q \in [l,n]} =  \frac{ | { {\bm j} \ra}  + | { {\bm p} \ra}  }{\sqrt{2}},
		\label{pluses}
	\end{eqnarray}
	where $| { {\bm j} \ra} = \bigotimes_i^q |j_i\ra$ and $| { {\bm p} \ra} = \bigotimes_i^q |p_i\ra$.  
	These states can be constructed in the same way as states (\ref{plus-l}). The missing elements can thus obtained 
	with, at most, $k$ measurements if performed element-wise.

	The full algorithm to compute $t^{k[n]}$ thus includes: 
	(i) At most $2^l - k \leq k$ expectation values via circuits of depth  $ \leq l+1$ to obtain $t^{k[l]}$ (yielding elements represented by open circles in  Fig.~\ref{fig:BellCirc});
	(ii) Filling the gap (pluses in Fig.~\ref{fig:BellCirc}) in $\mathds{1}_2 \otimes t^{k[m]}$ to produce $t^{k[m+1]}$, which requires measurements of at most $k$ projections on entangled states via circuits of depth at most $\left\lceil\log_2 2k \right\rceil$;
	(iii) A total of $n-l$ iterations to fill the entire band. 
	The total number of measurement circuits for $t^{k[n]}$  is thus
	\be
	\textrm{Comp}\left( t^{k[n]} \right) <
	\left(n + 1 - \log_2 k \right) k.
	\ee

	To construct the anti-diagonals $a^{k[n]}$, we first note that the expectation values of
	$a^{1[1]} = \begin{bmatrix}
		1 & 0 \\
		0 & 0
	\end{bmatrix}$, 
	$a^{2[1]} = \begin{bmatrix}
		0 & 1 \\
		1 & 0
	\end{bmatrix}$ and
	$a^{3[1]} = \begin{bmatrix}
		0 & 0 \\
		0 & 1
	\end{bmatrix}$ can be obtained by sampling one-qubit measurements of $\psi$ in $Z$ to get $\left\la a^{1[1]} \right\ra_\psi$ and $\left\la a^{3[1]} \right\ra_\psi$ and in $X$ to get $\left\la a^{2[1]} \right\ra_\psi$. 
	The elements $a^{k[m+1}]$ can be obtained by combining $\mathds{1}_2 \otimes a^{k[m]}$ and $a^{k[m]} \otimes \mathds{1}_2$, which  at most doubles the number of measurements \cite{sm}.
	
	
	We limit the construction to $r = {\log_2 \epsilon}/{\beta}  $ anti-diagonals.  Eqs. (\ref{off-diagonals1})-(\ref{off-diagonals2}) ensure that the remaining anti-diagonals will contribute less than $\epsilon$ to the expectation value (\ref{the-tau}).  
	The circuits can be constructed from $a^{k[1]}$ by incrementally increasing the number of qubits to  $a^{k[r]}$, which requires $\leq 2^r$ bases in total. The contributions from the anti-diagonals are obtained by measuring $n-r$ qubits in the computational basis and $a^{k[r]}$ thus constructed. 
	For the $n-s$ qubits, one needs to take into account only the all-$\uparrow$ and all-$\downarrow$ outputs, yielding the up-most and down-most $2^r$ anti-diagonals. The total number of bases for this protocol is bounded by
	\be
	2^{r} < 2^{1-\frac{\log_2 \epsilon}{\beta}} = 2\epsilon^{-\frac{1}{\beta}}.
	\ee

	\begin{table}[!htbp]
		\centering
		\tiny
		\begin{tabular*}{\columnwidth}{@{\extracolsep{\fill}}ccccccc}
			\toprule
			Electronic & \multirow{2}{*}{$v$} & \multirow{2}{*}{BM} & \multirow{2}{*}{$E_v$ (DVR)} & \multicolumn{3}{c}{$E_v^{\rm VQE}$} \\
			\cmidrule{5-7}
			state & & & & $\mathcal{C}_1$ & $\mathcal{C}_{0.01}$ & Linear \\
			\midrule
			\multirow{2}{*}{$^1\Sigma_g^+$}  & 0   & -15358.94 & -15358.99 & -15358.87 & -15358.99 & -15358.99 \\
			& 1   & -14846.67 & -14846.96 & -14838.70 & -14846.75 & -14846.96 \\
			& 2   & -14333.21 & -14332.81 & -14310.29 & -14332.80 & -14332.82 \\
			& 3   & -13826.93 & -13827.29 & -13797.65 & -13826.87 & -13827.29 \\
			& 4   & -13334.02 & -13335.45 & -13275.12 & -13318.77 & -13335.45 \\
			& 5   & -12861.37 & -12868.75 & -12897.70 & -12871.32 & -12868.75 \\
			\midrule
			\multirow{2}{*}{$^3\Sigma_u^+$}  & 0   & -9862.07 & -9862.14 & -9861.37 & -9862.14 & -9862.14 \\
			& 1   & -9559.46 & -9559.41 & -9538.97 & -9556.67 & -9559.41 \\
			& 2   & -9300.92 & -9300.88 & -9240.74 & -9266.82 & -9300.88 \\
			& 3   & -9080.60 & -9080.42 & 9068.09 & -9079.26 & -9085.40 \\
			& 4   & -8897.58 & -8896.82 & -8866.09 & -8870.57 & -8896.82 \\
			& 5   & -8747.53 & -8742.68 & -8729.41 & -8750.18 & ... \\
			\midrule
			\multirow{2}{*}{$^5\Sigma_g^+$}  & 0   & -7566.53 & -7566.50 &  -7565.88 & -7566.50 & -7566.50 \\
			& 1   & -7416.28 & -7416.29 &  -7397.28 & -7416.28 & -7416.29 \\
			& 2   & -7264.92 & -7264.69 &  -7181.98 & -7264.60 & -7264.69 \\
			& 3   & -7114.40 & -7118.43 &  -7075.88 & -7117.83 & -7118.43 \\
			& 4   & -6965.91 & -6958.08 &  -7001.98 & -6953.74 & -6958.08 \\
			& 5   & -6820.04 & -6840.02 &  -6873.37 & -6837.02 & -6840.02 \\
			\midrule
			\multirow{2}{*}{$^7\Sigma_u^+$}  & 0   & -6519.01 & -6519.04 & \multicolumn{2}{c}{-6519.04} & -6519.04 \\
			& 1   & -6350.36 & -6350.11 & \multicolumn{2}{c}{-6350.11} & -6350.11 \\
			& 2   & -6183.38 & -6185.17 & \multicolumn{2}{c}{-6185.17} & -6185.17 \\
			& 3   & -6018.09 & -6018.28 & \multicolumn{2}{c}{-6018.12} & -6018.28 \\
			& 4   & -5854.50 & -5849.10 & \multicolumn{2}{c}{-5848.69} & -5849.10 \\
			& 5   & -5692.63 & -5731.33 & \multicolumn{2}{c}{-5728.00} & -5731.33 \\
			\midrule
			\multirow{2}{*}{$^9\Sigma_g^+$}  & 0   & -5348.79 & -5348.82 & \multicolumn{2}{c}{-5348.82} & -5348.82 \\
			& 1   & -5175.81 & -5175.51 & \multicolumn{2}{c}{-5175.51} & -5175.51 \\
			& 2   & -5005.17 & -5008.56 & \multicolumn{2}{c}{-5008.55} & -5008.56 \\
			& 3   & -4836.85 & -4829.92 & \multicolumn{2}{c}{-4829.80} & -4829.92 \\
			& 4   & -4670.91 & -4683.68 & \multicolumn{2}{c}{-4683.42} & -4683.68 \\
			& 5   & -4507.31 & -4530.16 & \multicolumn{2}{c}{-4526.72} & -4612.83 \\
			\midrule
			\multirow{2}{*}{$^{11}\Sigma_u^+$} & 0   & -3677.68 & -3677.68 & -3677.00 & -3677.68 & -3677.68 \\
			& 1   & -3507.89 & -3507.82 & -3489.51 & -3507.82 & -3507.82 \\
			& 2   & -3341.77 & -3341.40 & -3253.24 & -3341.26 & -3341.40 \\
			& 3   & -3180.16 & -3186.93 & -3133.59 & -3185.51 & -3186.93 \\
			& 4   & -3023.07 & -3012.82 & -3061.61 & -3001.04 & -3012.82 \\
			& 5   & -2870.22 & -2874.49 & -2904.46 & -2866.05 & -2874.54 \\
			\midrule
			\multirow{2}{*}{$^{13}\Sigma_g^+$} & 0   & -548.68 & -548.68 & -548.65 & -548.67 & -548.68 \\
			& 1   & -497.16 & -497.15 & -496.47 & -496.84 & -497.15 \\
			& 2   & -449.18 & -449.26 & -443.48 & -448.36 & -449.26 \\
			& 3   & -404.71 & -404.67 & -382.88 & -390.96 & -404.67 \\
			& 4   & -363.58 & -362.99 & -369.09 & -360.62 & -362.99 \\
			& 5   & -325.67 & -325.72 & -315.46 & -310.29 & -325.72 \\
			\bottomrule
		\end{tabular*}
		\caption{Vibrational energy (in cm$^{-1}$) of \ch{Cr2} ($v =0-5$) in different electronic states. The benchmark (BM) results are obtained with a converged DVR basis. Exact energies obtained from the truncated DVR basis are displayed as $E_\nu$ (DVR). VQE uses quantum circuits shown in Fig. \ref{fig:cr2_pot_ansatz} (for $\mathcal{C}_1$) and Fig. \ref{fig:cr2_pot_ansatz2} (for $\mathcal{C}_{0.01}$). }
		\label{tab:diatomic}
	\end{table}

	\begin{figure*}
		\begin{tabular}{cc}
			\includegraphics[width=0.4\columnwidth]{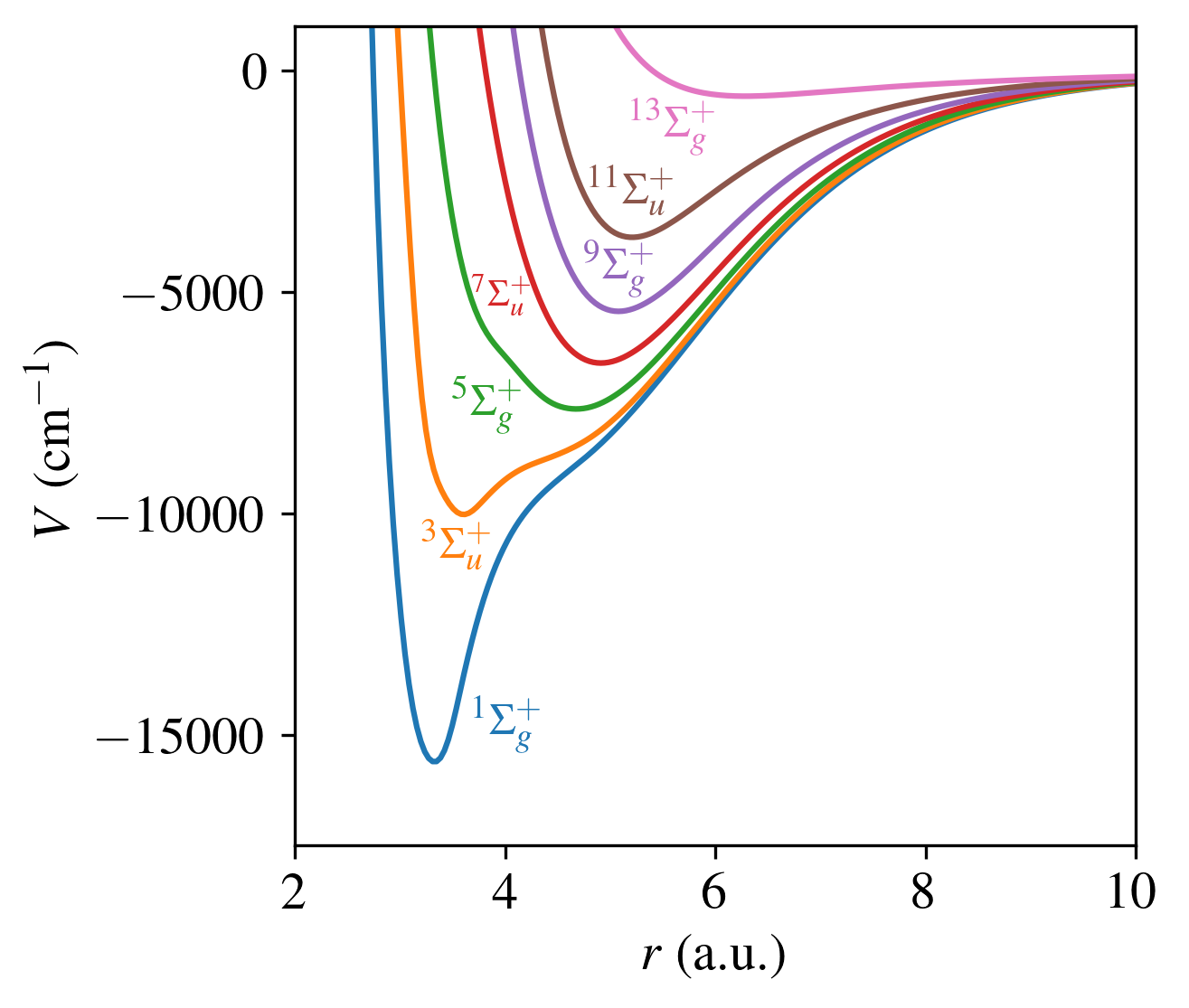} & 	\includegraphics[width=0.5\columnwidth]{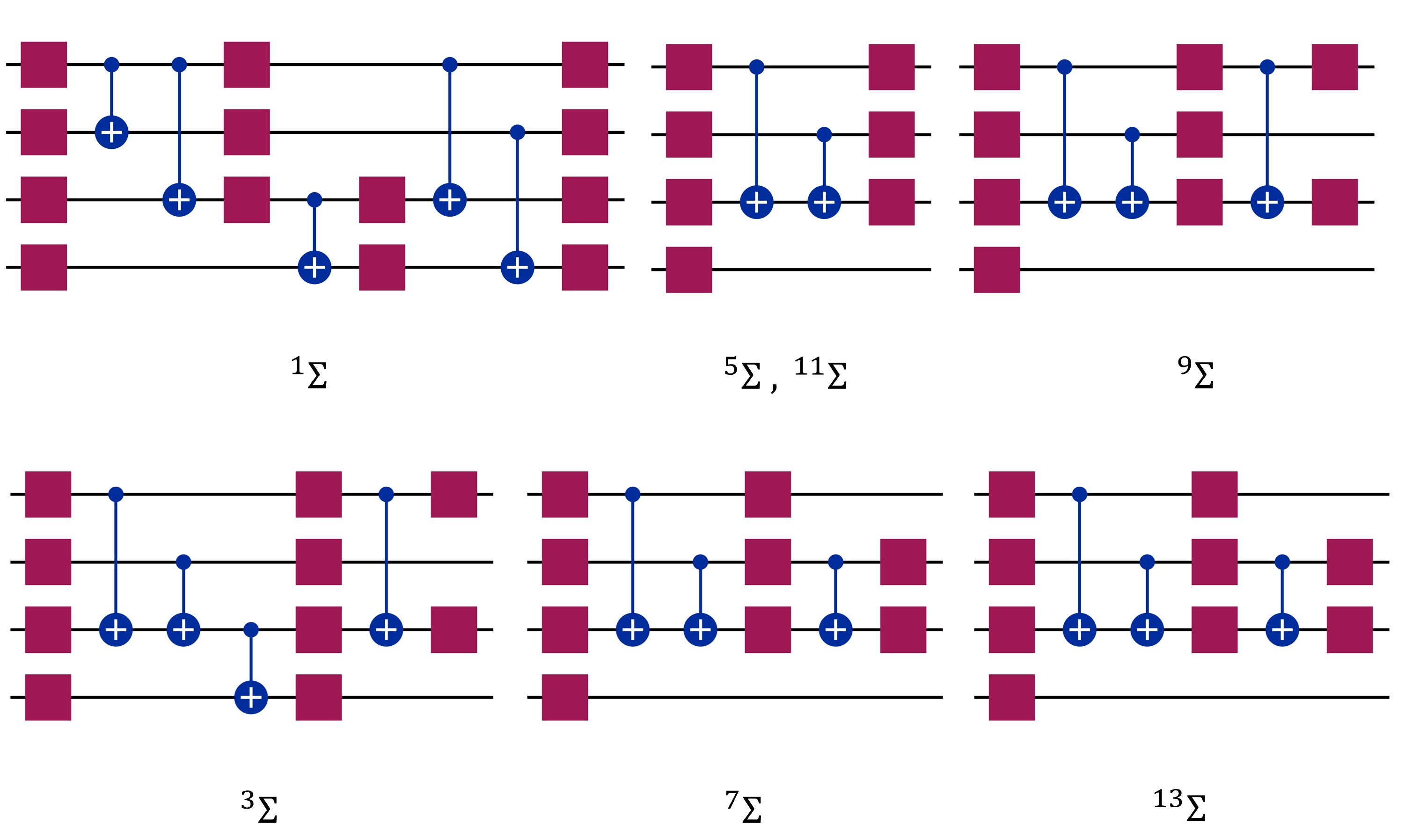} \\
		\end{tabular}
		\caption{Left: potential energy for Cr$_2$ from Ref. \cite{cr2-pot}. Right:  quantum circuits for VQE yielding the ground state energy with error $\leq 1$ cm${}^{-1}$. The squares represent the $R_Y$ gates and the circles show the entangling CNOT gates.}
		\label{fig:cr2_pot_ansatz}
	\end{figure*}

	\begin{figure*}
		\begin{tabular}{cc}
			\includegraphics[width=0.9\columnwidth]{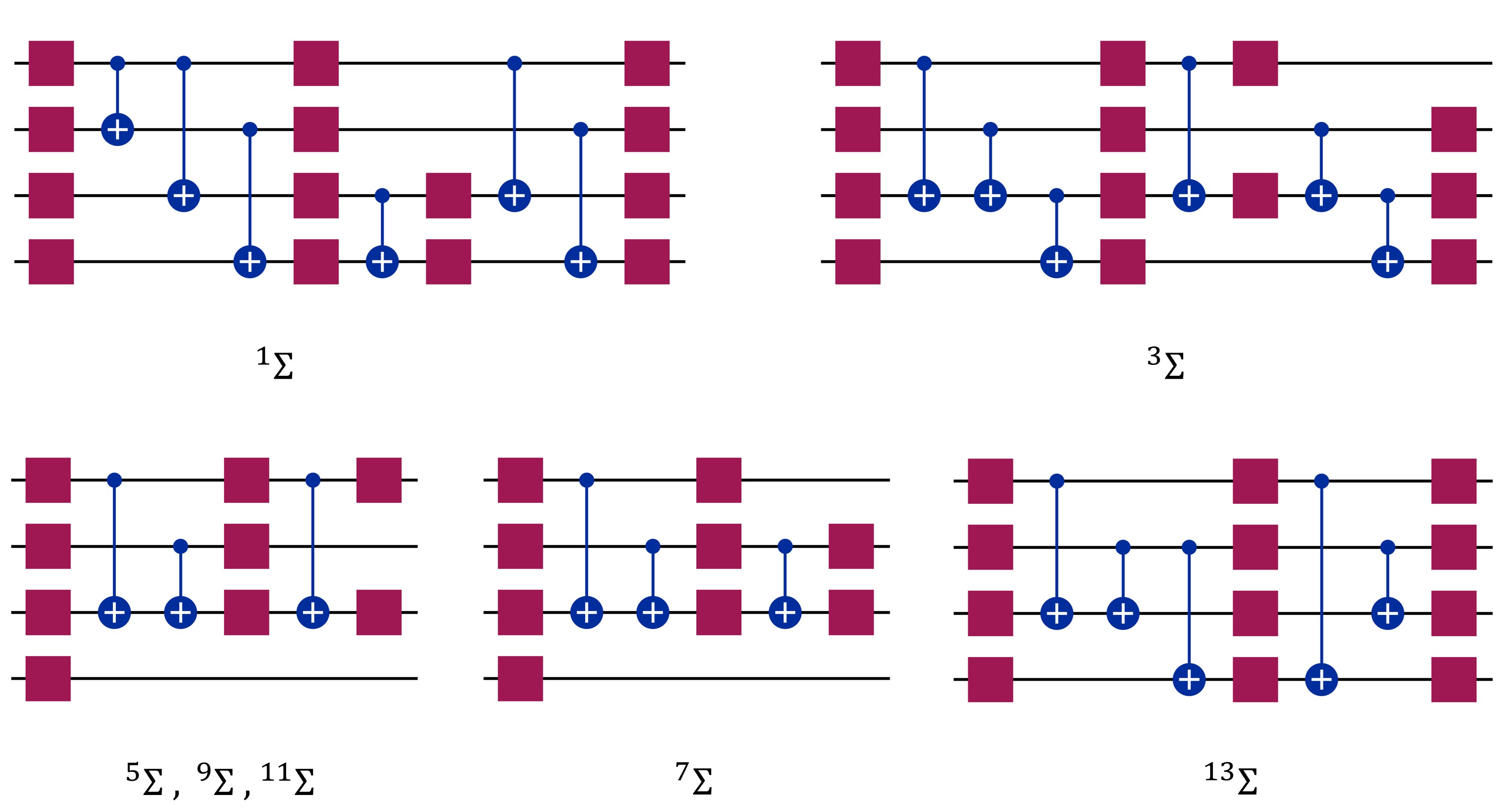} \\
		\end{tabular}
		\caption{Quantum circuits for VQE yielding the ground state energy with error $\leq 0.01$ cm${}^{-1}$. The squares represent the $R_Y$ gates and the circles show the entangling CNOT gates.}
		\label{fig:cr2_pot_ansatz2}
	\end{figure*}

	\subsection{Algorithm for compositional ansatz optimization}	
	
	\label{anzats-optimization} 
	
	In the second part of this work, we demonstrate that DVR Hamiltonians also lead to very efficient quantum circuits for representing vibrational states of molecules. 
	In order to apply VQE, it is necessary to find a proper ansatz for $| \psi \rangle$. It is not always clear how to select the ansatz for $| \psi \rangle$. 
	Previous work on electronic structure proposed various types of ansatzes for VQE with both fixed \cite{vqe-1,chem-3,uccsd,fixed-ansatz,uccgsd,symmetry,hva} and adaptive structure \cite{adaptvqe,adaptvqe-nuc,qubit-adaptvqe,iqcc,cluster-vqe,rotoselect,vans,evo-vqe,mog-vqe,qas}. 
	Unlike electronic structure problems, where interactions are pairwise additive, vibrational energy calculations are determined by a wide range of potential energy landscapes, which are highly molecule-specific. 
	In order to obtain the most efficient quantum circuit representations of $| \psi \rangle$ for VQE with DVR matrices, we develop and illustrate an iterative algorithm for ansatz construction that minimizes the number of entangling gates for each specific molecule. 
	This algorithm is inspired by work in Refs. \cite{comp-search-1, comp-search-2, xuyang}. 
	
	

	Our starting point is 
	\cite{chem-3}: 
	\begin{align}
		\ket{\psi(\bm\varphi)} = &\prod_{d=0}^{k-1}\left[\prod_{q=0}^{n-1} U^{q,d}(\varphi^q_{d})\times U^d_{\rm ent}\right] \nonumber\\
		&\times \prod_{q=0}^{N-1} U^{q, k}(\varphi^q_{k})\ket{0^n}, \label{eq:hea}
	\end{align}
	where $U^{q,d}(\varphi)$ represent $R_Y = \exp(-i\varphi\sigma_Y/2)$ for qubit $q$, and $k$ is the number of repetitions of the ansatz blocks. 
	The operator $U^d_{\rm ent}$ introduces entanglement between qubits. 
	The form of $U^d_{\rm ent}$ is determined by the ansatz construction algorithm that incrementally increases the complexity of the quantum circuits.  
	More specifically, 	
	the ansatz construction starts with a non-entangled quantum state given by Eq. (\ref{eq:hea}) with a predetermined number of blocks $k$  and $U^d_{\rm ent}$ set to identity. The method considers ${\rm CNOT}(q, p)~ \forall q < p$ as candidate gates for $U^d_{\rm ent}$, with each $d$ segment treated independently. In each optimization step, the candidate gate that lowers the VQE energy is added without replacement until convergence. 
	Here, we aim to converge the VQE calculation of the ground state either to 1 cm$^{-1}$ or 0.01 cm$^{-1}$, which yields quantum circuits of different complexity denoted $\mathcal{C}_1$ and $\mathcal{C}_{0.01}$. 
	This convergence error is with respect to the lowest eigenstate of the DVR matrix with the same number of DVR bases. 

	In order to benchmark the quantum circuits determined by this algorithm, we also use the following ansatz
	\begin{eqnarray}
		U_{\rm ent} = \prod_{q=0}^{n-1}{\rm CNOT}(q, q+1).
		\label{linear}
	\end{eqnarray}
	instead of $U^d_{\rm ent}$.
	The ansatz (\ref{linear}) is denoted hereafter as linear, as it linearly entangles adjacent qubits. 

	\begin{figure*}
			\includegraphics[width=\textwidth]{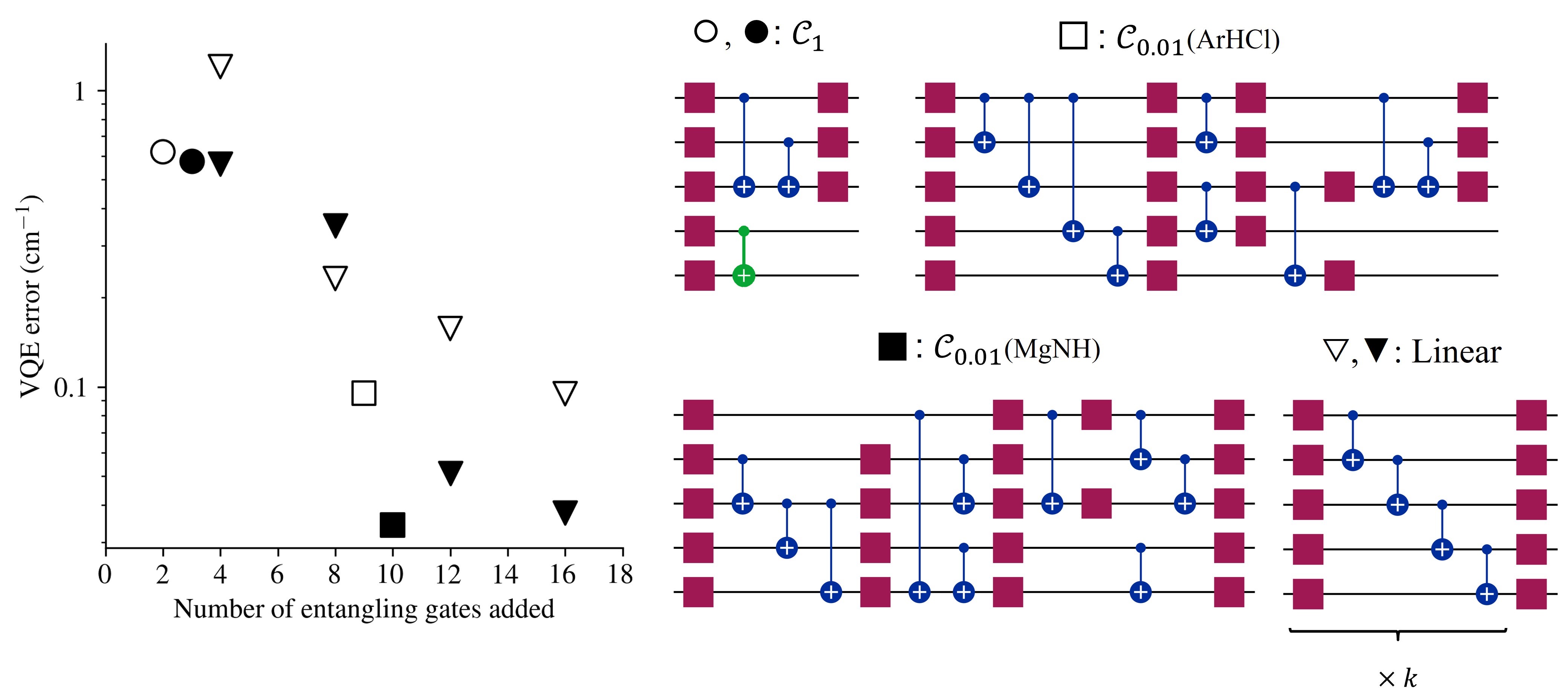}
		\caption{Left: VQE error for lowest energy of Ar--HCl (open symbols) and Mg--NH (full symbols) computed with optimized quantum circuits in the right panel. Circles -- with $\mathcal{C}_1$ circuits; squares -- with  $\mathcal{C}_{0.01}$ circuits; triangles -- with ansatz (\ref{eq:hea}) using (left to right) $k=1,2,3$ and $4$ and Eq. (\ref{linear}). 
			The $\mathcal{C}_1$ ansatz for Mg -- NH excludes the gate shown in green.}
		\label{fig:greedyent_const}
	\end{figure*}

	\begin{table*}
		\begin{tabular*}{\textwidth}{@{\extracolsep{\fill}}ccccccccc}
			\toprule
			\multirow{2}{*}{Molecule} & \multirow{2}{*}{$v$} & \multirow{2}{*}{\begin{tabular}{c}
					$E_v$ \\
					(experiment)
				\end{tabular}} & \multirow{2}{*}{\begin{tabular}{c}
					$E_v$ \\
					(computed in \cite{arhcl-pot})
				\end{tabular}} & \multirow{2}{*}{\begin{tabular}{c}
					$E_v$ \\
					(classical, present)
				\end{tabular}} & \multirow{2}{*}{\begin{tabular}{c}
					$E_v$ \\
					(DVR)
				\end{tabular}}  & \multicolumn{3}{c}{$E_{v}$ (VQE)} \\
			\cmidrule{7-9}
			& & & & & & $\mathcal{C}_1$ & $\mathcal{C}_{0.01}$ & Linear \\
			\midrule
			\multirow{3}{*}{ArHCl} & \multirow{1}{*}{0} & -114.7~\cite{arhcl-obs-0} & -115.151 & -115.265 & -115.178 & -114.645 & -115.169 & -115.171 \\
			& \multirow{1}{*}{1} & -91.04~\cite{arhcl-obs-1, arhcl-obs-12} & -91.485 & -91.642 & -90.959 & -80.824 & -90.485 & -90.929 \\
			& \multirow{1}{*}{2} & -82.26~\cite{arhcl-obs-2} & -82.717 & -82.825 & -82.727 & -75.900 & -82.986 & -82.650 \\
			\midrule
			\multirow{3}{*}{MgNH} & \multirow{1}{*}{0} & - & - & -88.227 & -88.196 & -87.650 & -88.191 & -88.190 \\
			& \multirow{1}{*}{1} & - & - & -63.603 & -62.928 & -56.050 & -62.730 & -62.664 \\
			& \multirow{1}{*}{2} & - & - & -55.461 & -54.961 & -55.145 & -54.850 & -54.866 \\
			\bottomrule
		\end{tabular*}
		\caption{Vibrational energy (in cm$^{-1}$) of Ar--HCl and Mg--NH by VQE with 32 DVR points and 5-qubit circuits. 
		}
		\label{tab:triatomic}
	\end{table*}

	\begin{figure*}
		\begin{tabular}{cc}
			\includegraphics[width=0.5\textwidth]{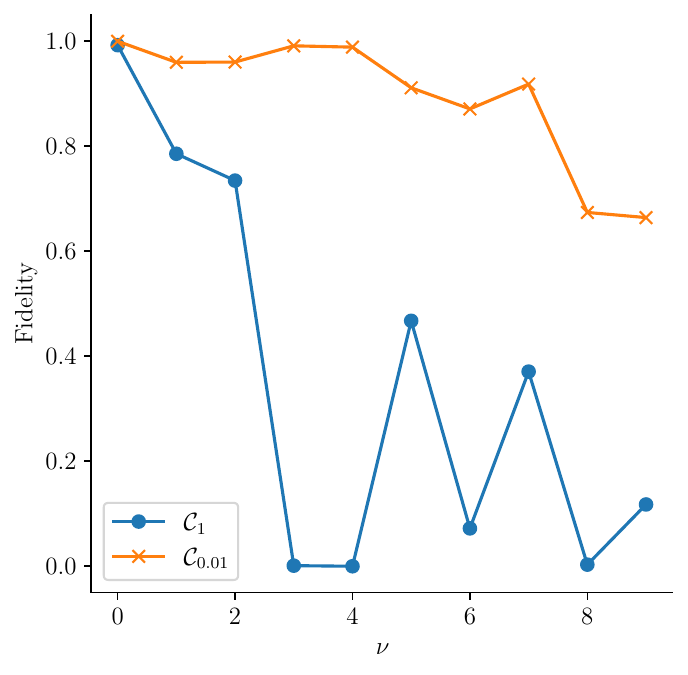} & 	\includegraphics[width=0.5\textwidth]{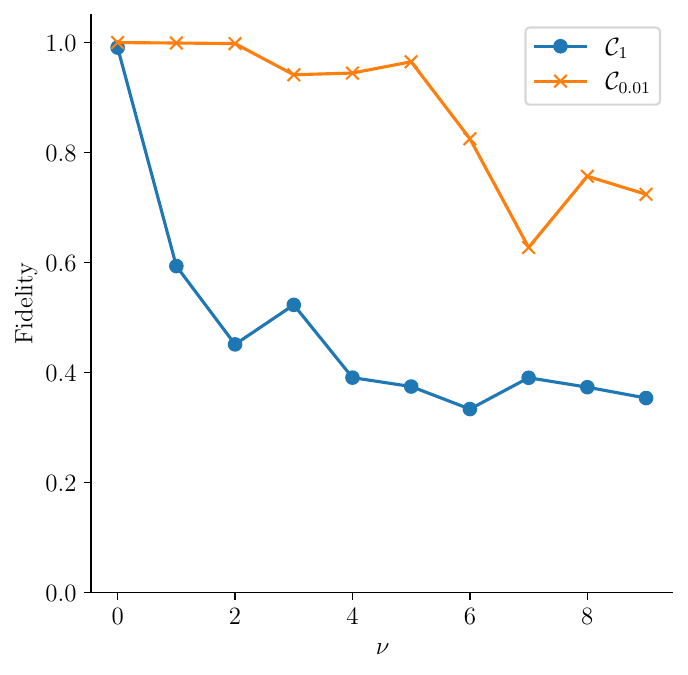}
		\end{tabular}
		\caption{Fidelity of the vibrational ground and excited states calculates with respect to the corresponding DVR wave functions for Ar--HCl (left) and Mg--NH (right).}
		\label{fig:fidelity}
	\end{figure*}

	\section{Numerical Results} 
	
	\label{numerical-results}
	
	We calculate the vibrational energy levels of diatomic (\ch{Cr2}) and triatomic (Ar--HCl and Mg--NH) systems. 
	We consider these molecular systems because they exhibit vibrational states with widely different energies (from $-55$ to $-15,000$ cm$^{-1}$ from the dissociation threshold) and spatial variations of wave functions and energy level patterns. Our goal is to demonstrate that the same approach can be applied to these widely different molecular systems. 
	For diatomic molecules we use the DVR Hamiltonian from Ref. \cite{colbert}. For triatomic complexes, we use the DVR approach by Choi and Light \cite{choi}. 
	We use two classical constrained optimization methods to optimize quantum circuit parameters: 
	the bounded limited memory method of  Broyden, Fletcher, Goldfarb, and Shanno 
	\cite{lbfgsb-1, lbfgsb-2};
	and sequential least squares programming \cite{slsqp}. 
	We benchmark the VQE results by the vibrational levels calculated using direct diagonalization with the converged DVR basis and previous literature results, where available.  
	
	Table \ref{tab:diatomic} demonstrates the performance of VQE for vibrational states $v=0$ -- $5$ of seven electronic states 
	of Cr$_2$ with zero rotational angular momentum. We use the interaction potentials from Ref. \cite{cr2-pot}, illustrated in Fig. \ref{fig:cr2_pot_ansatz}.  The VQE calculations use 16 DVR points placed to span the range including
	the minimum of the potential energy. Apart from the DVR parameters, the circuit optimization algorithm is applied identically to all seven electronic states. The Hamiltonian is represented by 
	the expansion (\ref{Pauli}) including $\approx 130$ Pauli terms. 
	Table \ref{tab:diatomic} displays VQE results obtained with three types of quantum circuits:  the linear ansatz (\ref{linear}) with 3 repetitions, and optimized circuits ${\cal C}_1$ (shown in Fig. \ref{fig:cr2_pot_ansatz}) and ${\cal C}_{0.01}$ (shown in Fig. \ref{fig:cr2_pot_ansatz2}). 
	
	The results in Table  \ref{tab:diatomic} and Figs. \ref{fig:cr2_pot_ansatz} and \ref{fig:cr2_pot_ansatz2} show that VQE can be used to compute the vibrational levels of diatomic molecules with high precision using shallow quantum circuits. It is particularly instructive to analyze the difference between the quantum circuits displayed in Figs. \ref{fig:cr2_pot_ansatz} and \ref{fig:cr2_pot_ansatz2}. The quantum circuits in Fig. \ref{fig:cr2_pot_ansatz2} yield a much better convergence (within 0.01 cm$^{-1}$) and a significantly higher accuracy than the quantum circuits in Fig. \ref{fig:cr2_pot_ansatz} at the expense of a small number of additional quantum gates. For example for the ground vibrational state of Cr$_2$ in the $^{3}\Sigma_g^+$ electronic state, adding two entanglement gates reduces the error of the computation from 0.69 cm$^{-1}$ to $0.07$ cm$^{-1}$, with respect to the benchmark calculation result (labeled BM). 
	
	These results also illustrate the utility of the ansatz optimization algorithm developed here. By construction, the algorithm aims to produce quantum circuits of the lowest complexity for a particular accuracy target or particular convergence threshold. This can be exploited to examine the role of specific gates or gate combinations in determining the expressivity of quantum circuits for representing the vibrational states of molecules. Given that the optimized quantum circuits illustrated in Figs. \ref{fig:cr2_pot_ansatz} and \ref{fig:cr2_pot_ansatz2}  are very shallow, 
	the results of Table I suggest that current quantum computing devices can already be used for high-precision computation of vibrational energy levels.

	For tri-atomic complexes, we use 32 DVR basis states and accurate atom - molecule potential energy surfaces by Hutson for Ar--HCl \cite{arhcl-pot} and by Sold\'an et al. for Mg--NH \cite{mgnh-pot}. We keep both HCl and NH in the ground vibrational state and compute the vibrational states supported by the atom - molecule interaction potential.  We obtain the DVR points for the triatomic systems by diagonalizing the coordinate representations. 
	The parameters to generate the DVR points are selected to cover the low-energy regions of the potential energy surface. 
	The DVR Hamiltonians are represented by 165 (Ar--HCl) and 170 (Mg--NH) Pauli terms in Eq. (\ref{Pauli}) acting on 5 qubits.
	As above, we construct three types of quantum circuits: ${\cal C}_1$,  ${\cal C}_{0.01}$ and the linear ansatz (\ref{linear}) with $k$ repetitions.

	We note that the triatomic molecular systems consider here represent van der Waals complexes. 
	Accurate calculations of ro-vibrational energies for van der Waals complexes are generally challenging due to the weak binding of the underlying potential energy surface and a large spatial extent of the corresponding vibrational states. Typically, ro-vibrational energy levels of van der Waals complexes are computed by solving a system of coupled differential equations, using, for example, a log-derivative propagation method. 
	Figure \ref{fig:greedyent_const}
	and Table \ref{tab:triatomic} show that accurate results for these weakly bound molecular complexes can be obtained by VQE with very shallow circuits. Figure \ref{fig:fidelity} shows the fidelity of the ground and excited states obtained from $\mathcal{C}_1$ and $\mathcal{C}_{0.01}$, with respect to the corresponding DVR eigenstates and demonstrates that the quantum circuits not only provide accurate energies, but also a good representation of the underlying eigenstates. 
	We observe that a small number of entangling gates is sufficient to ensure accurate calculations for both the ground  and excited state energy. 
	We also observe that some of the optimized circuits in Figs. \ref{fig:cr2_pot_ansatz} and \ref{fig:greedyent_const} are only partially entangled, which indicates that 
	accurate VQE results can be obtained with ensembles of unentangled quantum circuits. This further decreases the complexity of the quantum computations.  
	
	\begin{figure*}
		\begin{tabular}{cc}
			\multicolumn{2}{c}{\includegraphics[width=\textwidth]{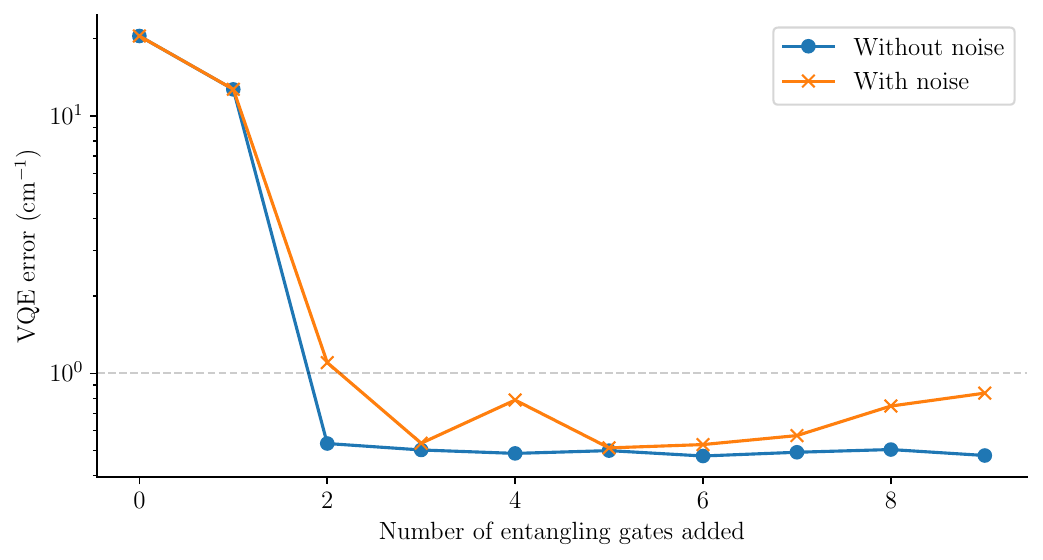}} \\
			\includegraphics[width=0.5\textwidth]{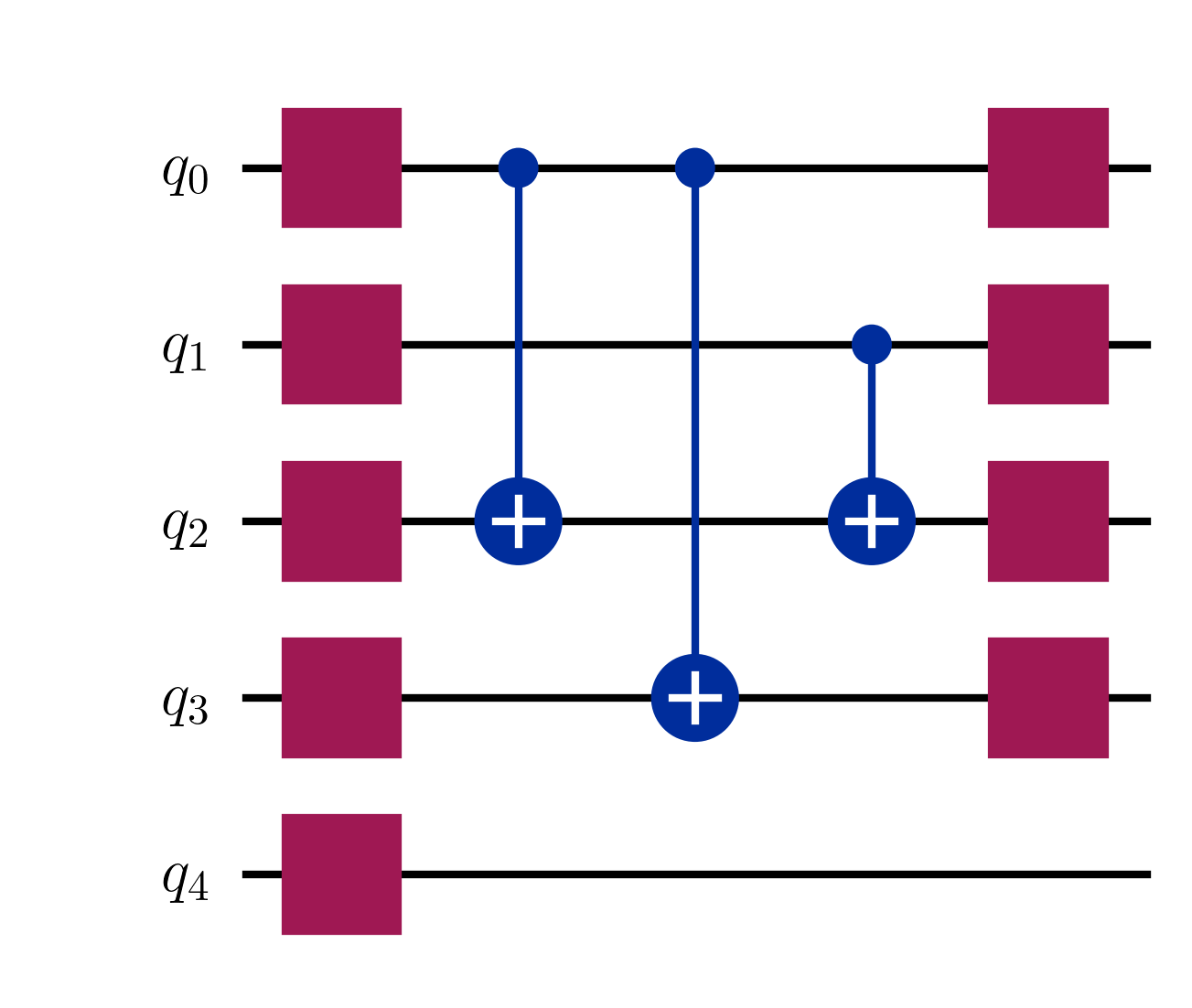} & \includegraphics[width=0.43\textwidth]{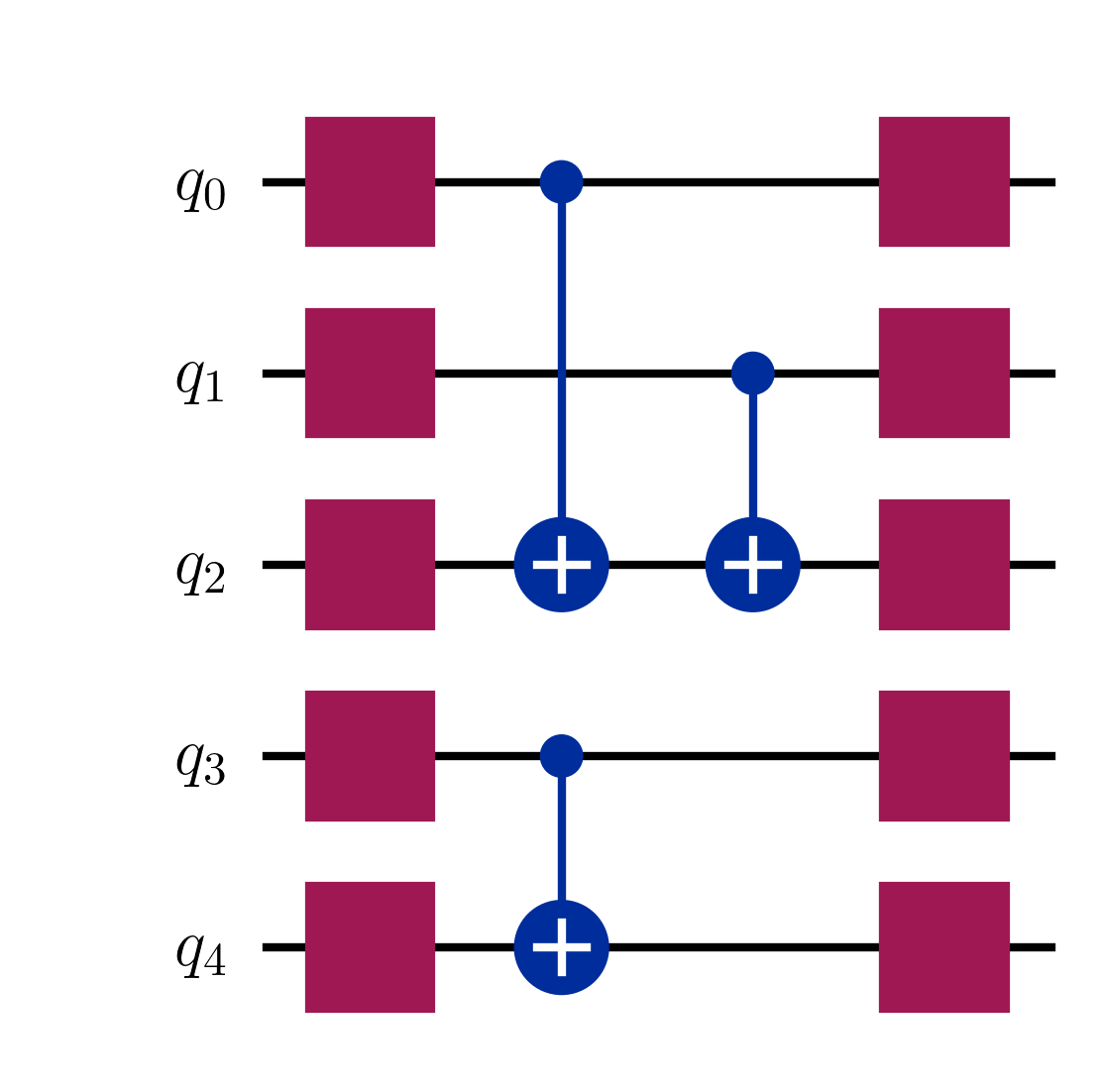} \\
			\quad\qquad With noise & \quad\quad\qquad Without noise
		\end{tabular}
		\caption{(Top) VQE error during the greedy circuit optimization for Ar--HCl using $k=1$ in Eq.~\eqref{eq:hea}. (Bottom) The circuits obtained after adding 3 entangling gates with and without the noise model for Ar--HCl.}
		\label{fig:noisy_opt}
	\end{figure*}
	
	In order to assess whether the present method is practical for NISQ-era devices, we also ran the circuit optimization algorithm for the Ar--HCl complex with a realistic noise model. Specifically, we repeat the full compositional ansatz search using the Sherbrooke noise model in Qiskit, which captures gate and readout errors of a superconducting quantum processor. Figure \ref{fig:noisy_opt} shows the error of the ground state VQE calculation during the circuit optimization algorithm for circuits with $k=1$ in Eq.~\eqref{eq:hea}, and two circuits obtained after 4 steps of the greedy search algorithm. We find that the greedy optimization continues to identify shallow circuits yielding spectroscopic accuracy for multiple excited vibrational states. While the circuits obtained under noise may differ from those found in noiseless simulations, their accuracy is comparable, indicating the existence of multiple noise-resilient shallow representations of the same vibrational eigenstates. The persistence of successful optimization under realistic noise and the existence of multiple shallow optima suggest that the DVR-based ansatz space avoids the effects typically associated with barren plateaus in deep, unstructured circuits. Further analysis of the robustness of the method to noise is left for future work. 
	
	\section{Conclusions}	
	
	\label{summary}

	We have shown that the structure of DVR matrices can be leveraged 
	to represent molecular Hamiltonians by the number of quantum circuits that grows polynomially with the number $n$ of qubits. Given the binary basis encoding yielding $N = 2^n$, this offers an exponential reduction of the measurement complexity for matrices with $N$ DVR basis states. 
	This has significant implications for the potential quantum advantage of 
	VQEs for ro-vibrational computations. Exponential scaling of VQE is a significant and common challenge, particularly relevant for unstructured Hamiltonians such as the vibrational eigenvalue problem in an unstructured basis. For quantum chemistry calculations as well as for ro-vibrational problems, this is usually dealt with by second quantization. However, these approaches often lead to extended quantum circuits that require a large number of qubits and gates.

	We have also demonstrated that DVR leads to efficient quantum circuits for VQE computations of vibrational energy levels. 
	To show this, we have introduced a general approach to constructing the quantum ansatz by combining DVR with VQE and a greedy search in the space of gate permutations.  The results yield compact representations of vibrational states by quantum circuits of a gate-based quantum computer.  
	We have shown that both the ground and excited vibrational energies can be computed with the relative accuracy of $< 1 \%$ using very simple, in some cases, partially entangled circuits. The accuracy of $1$ cm$^{-1}$ can be achieved with $< 20$ ($< 5$ entangling) gates and 4 qubits for diatomic molecules and $< 30$ ($< 9$ entangling) gates with 5 qubits for triatomic van der Waals complexes. 
	This should be compared with previous VQE calculations of vibrational energy levels that required extended quantum circuits with $>200$ (for CO, COH and O$_3$ molecules  \cite{mol-vib-3}) or between 44 and 140 292 (for CO$_2$, H$_2$CO and HCOOH molecules \cite{mol-vib-2}) entangling gates. 
	
	Reducing the complexity of quantum circuits is particularly important for VQE approaches that are known to be affected by barren plateaus, which limit the quantum advantage of variational quantum computations. Although barren plateaus are formally an asymptotic effect, they can become relevant already for systems of moderate size. While a comprehensive treatment of barren plateaus is beyond the scope of the present work, our results provide several encouraging indications that this issue may be mitigated for vibrational energy calculations using the present framework. In particular, we observe no signs of barren plateaus for the systems considered here, including in simulations that explicitly incorporate realistic noise. Moreover, the anzatz construction algorithm introduced in this work consistently yields accurate vibrational energies and wave functions using extremely shallow circuits, some of which are only partially entangled. These observations suggest that VQE computations of vibrational energies may be scalable to larger systems than can be anticipated based on the previous calculations\cite{mol-vib-2, mol-vib-3}.  This must be confirmed by a more systematic numerical investigation, which we leave for future work.
	
	Finally, we note that DVR does not require global fits of potential energy surfaces or integrals over the potential energy for the construction of the Hamiltonian matrix. 
	The implementation of VQE with DVR in first quantization does not require normal-mode analysis for the construction of the quantum ansatz. 
	For these reasons, the present approach can be readily extended beyond molecular dynamics. For example, this method can be directly applied to designing efficient quantum circuits for variational computations of the eigenspectra of semiconductor quantum dots \cite{demler}. Finally, our calculations suggest that quantum circuit representations of vibrationals states can be used as efficient (small number of free parameters) ansatze for classical variational calculations. 
	
	\section{Author contributions}
	All authors have designed the concept of this work
	and co-wrote the manuscript. The algorithm presented
	in Section IIA was developed by DB. The
	algorithm for ansatz optimization was developed by
	KA and RVK and the numerical calculations were
	performed by KA.
	
	\section{Conflicts of interest}
	
	There are no conflicts to declare.
	
	\section{Data Availability}
	The data that support the findings of this study
	are available within the article and its supplementary
	material.
	
	\appendix	
	
	\section{Scaling of DVR matrix bands}

	The goal of this Appendix is to prove Eqs. (\ref{off-diagonals1}) and (\ref{off-diagonals2}). Specifically, we aim to show that
	\be
	\sum_{k=s}^{2^n-1}\left| f(k) \right| \leq O\left( s^{-\alpha} \right), \ \alpha>0, \label{Prop_f}\\
	\sum_{k=r}^{2^{n+1}-1-r}\left| g(k) \right| \leq O\left(  r^{-\beta} \right), \ \beta>0, \label{Prop_g}
	\ee
	where $r \ll N$ and $s \ll N$, and $N = 2^n$ is assumed to be large.  
	
	We consider a particle with mass $m$ on a grid of $x=[a,b]$ with the uniform spacing $\Delta x$, yielding
	\be
	x_j = b + j\Delta x, \quad j = 1,\dots, N-1.
	\ee
	Given an orthonormal basis $\left\{\phi_n \right\}$, the matrix elements of the kinetic energy  
	can be written as
	\be
	T_{ij} =   - \frac {\hbar^2}{2m} \Delta x \sum_{n=0}^{N-1}
	\phi_n (x_i) \phi_n^{''} (x_j).
	\ee
	Fourier grid DVR uses the basis  \cite{colbert}
	\be
	\phi_n(x) = \left(\frac{2}{b-a}\right)^{1/2} 
	\sin\left[ \frac{n \pi (x-a)}{b-a} \right].
	\ee
	For this basis, the sum can be evaluated analytically, yielding \cite{colbert}
	\be \label{KEDVR}
	T_{ij} = 
	\begin{cases}
		\frac {\hbar^2}{2m} \frac{1}{(b-a)^2}\frac{\pi^2}{2}\left[ \frac{2 N^2 +1}{3} - \frac{1}{\sin^2 (\pi j / N) }\right], & \ i=j,\\
		\frac {\hbar^2}{2m} \frac{(-1)^{i-j}}{(b-a)^2}\frac{\pi^2}{2}
		\left\{  \frac{1}{\sin^2 [\pi (i-j) / 2 N] } -  \frac{1}{\sin^2 [\pi (i+j) / 2 N] }\right\}, & \ i \neq j.
	\end{cases}
	\ee
	%
	
	Following Colbert and Miller \cite{colbert}, we now consider the cases of infinite lattices, relevant for radial molecular coordinates, and a case of finite $a$ and $b$.

	\subsection{Infinite lattice $[a= -\infty,b = \infty ]$}
	\label{infinite}
	
	The finite grid spacing $\Delta x = \frac{b-a}{N}$ requires $N \rightarrow \infty$. In this limit, Eq.~(\ref{KEDVR}) becomes
	\be \label{infDVR}
	T_{ij} = \frac {\hbar^2}{2m \Delta x^2}(-1)^{i-j}
	\begin{cases}
		\frac {\pi^2}{3},  & \ i=j,\\
		\frac {2}{(i-j)^2}, & i \neq j.
	\end{cases}
	\ee
	Note that in this case $g(i+j)$ vanishes. It is evident that 
	\be
	| f(k)| =  \frac {2}{(i-j)^2}
	\ee
	satisfies Eq.~(\ref{Prop_f}). To prove this, we note that for functions that are (strictly) monotonically decaying in absolute value
	\be \label{ineq:int}
	\int_{s}^{u} |f(k)| dk < \sum_{k=s}^{u} |f(k)| < \int_{s-1}^{u-1} |f(k)| dk.
	\ee
	For our choice of $f(\cdot)$ this yields the bound
	\be \label{f:proof}
	\sum_{k=s}^{2^n - 1} |f(k)| < 2  \int_{s-1}^{2^n - 2} \frac{dk}{k^2} = 2\left( \frac{1}{s-1} - \frac{1}{2^n-2} \right) < 3  s^{-1}.
	\ee

	\subsection{Infinite lattice $[a= 0,b = \infty ]$} \label{ss:halfinfinite}
	
	For the lattice with $[a= 0,b = \infty ]$, the matrix elements
	\be
	T_{ij} = \frac {\hbar^2}{2m \Delta x^2}(-1)^{i-j}
	\begin{cases}
		\frac {\pi^2}{3} - \frac{1}{2i^2},  & \ i=j,\\
		\frac {2}{(i-j)^2} - \frac {2}{(i+j)^2}, & i \neq j.
	\end{cases}
	\ee
	include $g(k)$. Since Eq. (\ref{f:proof}) applies to $g(k)$, the arguments in Subsection (\ref{infinite}) apply to Eq. (12).

	
	\subsection{Finite lattice $[a,b]$}
	We note that the prefactor in Eq.~(\ref{KEDVR}) satisfies $\frac {\hbar^2}{2m} \frac{1}{(b-a)^2}\frac{\pi^2}{2} \cdot \frac{2m \Delta x^2}{\hbar^2} = \frac{\pi^2 }{2 N^2}$. For the following functions
	\be
	|\tilde f(k)| &=& \frac{\pi^2 }{2 N^2}
	\left[  \frac{1}{\sin^2 (\pi k / 2 N) } -  1\right],\\
	|\tilde g(k)| &=& \frac{\pi^2 }{2 N^2}
	\left[ \frac{1}{\sin^2 (\pi k / 2 N) } - 1\right],
	\ee
	Eq.~(\ref{ineq:int}) yields
	\be
	\sum_{k=s}^{N-1}  \frac{1}{\sin^2 (\pi k / 2 N) } &=& \frac{2 N}{\pi} \sum_{k=s} \frac{\pi}{2 N} \frac{1}{\sin^2 (\pi k / 2 N) }\\
	&<& \frac{2 N}{\pi} \int_{\frac{\pi(s-1)}{2 N}}^{\frac{\pi}{2}} \frac{dx}{\sin^2(x)}\\ \label{sum:int}
	&=& \frac{2 N}{\pi} \cot \left[ \frac{\pi(s-1)}{2 N} \right].
	\ee
	Using the Laurent series for $\cot(y)$ around $y=0$, we can write
	\be
	\frac{2 N}{\pi} \cot \left[ \frac{\pi(s-1)}{2 N} \right] = \frac{4 N^2}{\pi^2 (s-1)} - \frac{(s-1)}{3} + O\left(\frac{s^3}{N^2}\right).
	\ee
	This yields 
	\be
	\sum_{k=s}^{2^n-1} |\tilde f(k)| &<&  \frac{\pi^2 }{2 N^2}
	\left[ \frac{4 N^2}{\pi^2 (s-1)} - \frac{(s-1)}{3} -  (N-1-s)  + O\left(\frac{s^3}{N^2}\right)\right]\\
	\label{1}
	&<&  \frac{2}{s-1} + O\left(\frac{s^3}{N^4}\right) \\
	&<& 3 s^{-1}.
	\label{2}
	\ee
	
	For $g(k)$, we exploit 
	the symmetry of $\sin(\cdot)$ to write
	\be
	\sum_{k=r}^{2^{n+1}-1-r} \frac{1}{\sin^2 (\pi k / 2 N) } = 1 + 2 \sum_{k=r}^{2^{n}-1} \frac{1}{\sin^2 (\pi k / 2 N) },
	\label{3}
	\ee
	and apply Eqs. (\ref{1})-(\ref{3}).


	\section{Representation of $H^{(s,r)}$ in terms of $t^{k[n]}$ and $a^{k[n]}$}
	
	We leverage the structure of the truncated DRV matrix by representing it as a sum of a diagonal matrix $D$ and $k^{\text{th}}$ diagonal and anti-diagonal components,  
	\be 
	H^{\left(s,r\right)}
	&=& D + \sum_{k=1}^{s-1} f(k) t^{k[n]} + \sum_{k=1}^{r-1} \left( g(k) a^{k[n]} + g(2^n-k) a^{(2^n-k)[n]} \right),
	\ee
	where
	\be
	D_{ij} &=& \delta_{ij}\left( d(i) - g(2i) - \sum_{k=1}^{s-1}f(k) q^{k[n]}(i)  \right)\\
	t^{k[n]}_{ij} &=& \delta_{|i-j|, k} + \delta_{ij}q^{k[n]}(i),\\
	a^{k[n]}_{ij} &=& \delta_{i+j, k}   
	\ee
	and $q^{k[n]}(i)$ can be computed by an efficient Algorithm 1. Our strategy is to measure the expectation values of each of the $\left\{D, t^{k[n]}, a^{k[n]} \right\}$ separately using the induction over $n$ and sum them with the corresponding weights.
	
	In addition to $k^{\text{th}}$ diagonals, the construction of $t^{k[n]}$ adopted in this work introduces additional contributions  $q^{k[n]}(i)$ to the matrix elements on the main diagonal. These contributions can be deduced from the measurement protocol for $t^{k[n]}$ as described in  Algorithm~\ref{alg:Q}.
	\begin{algorithm}[H]
		\caption{Computing $q^{k[n]}(i)$ from binary representation of $i$}\label{alg:Q}
		\begin{algorithmic}
			\State \textbf{define} $l \gets \lceil \log_2 (k+1) \rceil$,
			\State $\quad last\_l\_bits \gets i \mod 2^{l}$,
			\State $\quad anti\_last\_l\_bits \gets 2^{l} - last\_l\_bits$;
			\State \textbf{initialize} $output \gets 1$,
			\State $\quad j \gets l+1$;
			\While {$j \leq n$}
			\If {$j^{\mathrm{th}}$ bit of $i$ is 1}
			\If {$last\_l\_bits < k$ or $anti\_last\_l\_bits < l$} 
			\State $output \rightarrow out+1$;
			\EndIf
			\EndIf
			\State $j \rightarrow j+1$;
			\EndWhile
			\Return $output$.
		\end{algorithmic}
	\end{algorithm}

	\acknowledgments 
	
	This work was supported by NSERC of Canada and the Stewart Blusson Quantum Matter Institute. 	
	
	\bibliography{manuscript}

@article{lbfgsb-1,
	author = {Byrd, Richard H. and Lu, Peihuang and Nocedal, Jorge and Zhu, Ciyou},
	title = {A Limited Memory Algorithm for Bound Constrained Optimization},
	journal = {SIAM J. Sci. Comput.},
	volume = {16},
	number = {5},
	pages = {1190-1208},
	year = {1995},
	doi = {10.1137/0916069},
	
	URL = { 
	https://doi.org/10.1137/0916069
	
	}
}

@article{lbfgsb-2,
	author = {Zhu, Ciyou and Byrd, Richard H. and Lu, Peihuang and Nocedal, Jorge},
	title = {Algorithm 778: {L-BFGS-B}: Fortran Subroutines for Large-Scale Bound-Constrained Optimization},
	year = {1997},
	issue_date = {Dec. 1997},
	publisher = {Association for Computing Machinery},
	address = {New York, NY, USA},
	volume = {23},
	number = {4},
	issn = {0098-3500},
	url = {https://doi.org/10.1145/279232.279236},
	doi = {10.1145/279232.279236},
	journal = {ACM Trans. Math. Softw.},
	month = {dec},
	pages = {550–560},
	numpages = {11}
}

@book{slsqp,
	title={A Software Package for Sequential Quadratic Programming},
	author={Kraft, D.},
	series={Deutsche Forschungs- und Versuchsanstalt f{\"u}r Luft- und Raumfahrt K{\"o}ln: Forschungsbericht},
	url={https://books.google.ca/books?id=4rKaGwAACAAJ},
	year={1988},
	publisher={Wiss. Berichtswesen d. DFVLR}
}

@Article{vqe-1,
	author={Peruzzo, Alberto
	and McClean, Jarrod
	and Shadbolt, Peter
	and Yung, Man-Hong
	and Zhou, Xiao-Qi
	and Love, Peter J.
	and Aspuru-Guzik, Al{\'a}n
	and O'Brien, Jeremy L.},
	title={A variational eigenvalue solver on a photonic quantum processor},
	journal={Nat. Commun.},
	year={2014},
	month={Jul},
	day={23},
	volume={5},
	number={1},
	pages={4213},
	issn={2041-1723},
	doi={10.1038/ncomms5213},
	url={https://doi.org/10.1038/ncomms5213}
}

@article{vqe-2,
	doi = {10.1088/1367-2630/18/2/023023},
	url = {https://dx.doi.org/10.1088/1367-2630/18/2/023023},
	year = {2016},
	month = {feb},
	publisher = {IOP Publishing},
	volume = {18},
	number = {2},
	pages = {023023},
	author = {Jarrod R McClean and Jonathan Romero and Ryan Babbush and Alán Aspuru-Guzik},
	title = {The theory of variational hybrid quantum-classical algorithms},
	journal = {New J. Phys.}
}

@article{chem-1,
	title = {Scalable Quantum Simulation of Molecular Energies},
	author = {O'Malley, P. J. J. and Babbush, R. and Kivlichan, I. D. and Romero, J. and McClean, J. R. and Barends, R. and Kelly, J. and Roushan, P. and Tranter, A. and Ding, N. and Campbell, B. and Chen, Y. and Chen, Z. and Chiaro, B. and Dunsworth, A. and Fowler, A. G. and Jeffrey, E. and Lucero, E. and Megrant, A. and Mutus, J. Y. and Neeley, M. and Neill, C. and Quintana, C. and Sank, D. and Vainsencher, A. and Wenner, J. and White, T. C. and Coveney, P. V. and Love, P. J. and Neven, H. and Aspuru-Guzik, A. and Martinis, J. M.},
	journal = {Phys. Rev. X},
	volume = {6},
	issue = {3},
	pages = {031007},
	numpages = {13},
	year = {2016},
	month = {Jul},
	publisher = {American Physical Society},
	doi = {10.1103/PhysRevX.6.031007},
	url = {https://link.aps.org/doi/10.1103/PhysRevX.6.031007}
}

@article{chem-2,
	title = {Quantum implementation of the unitary coupled cluster for simulating molecular electronic structure},
	author = {Shen, Yangchao and Zhang, Xiang and Zhang, Shuaining and Zhang, Jing-Ning and Yung, Man-Hong and Kim, Kihwan},
	journal = {Phys. Rev. A},
	volume = {95},
	issue = {2},
	pages = {020501},
	numpages = {5},
	year = {2017},
	month = {Feb},
	publisher = {American Physical Society},
	doi = {10.1103/PhysRevA.95.020501},
	url = {https://link.aps.org/doi/10.1103/PhysRevA.95.020501}
}

@Article{chem-3,
	author={Kandala, Abhinav
	and Mezzacapo, Antonio
	and Temme, Kristan
	and Takita, Maika
	and Brink, Markus
	and Chow, Jerry M.
	and Gambetta, Jay M.},
	title={Hardware-efficient variational quantum eigensolver for small molecules and quantum magnets},
	journal={Nature},
	year={2017},
	month={Sep},
	day={01},
	volume={549},
	number={7671},
	pages={242-246},
	issn={1476-4687},
	doi={10.1038/nature23879},
	url={https://doi.org/10.1038/nature23879}
}

@article{chem-4,
	title = {Computation of Molecular Spectra on a Quantum Processor with an Error-Resilient Algorithm},
	author = {Colless, J. I. and Ramasesh, V. V. and Dahlen, D. and Blok, M. S. and Kimchi-Schwartz, M. E. and McClean, J. R. and Carter, J. and de Jong, W. A. and Siddiqi, I.},
	journal = {Phys. Rev. X},
	volume = {8},
	issue = {1},
	pages = {011021},
	numpages = {7},
	year = {2018},
	month = {Feb},
	publisher = {American Physical Society},
	doi = {10.1103/PhysRevX.8.011021},
	url = {https://link.aps.org/doi/10.1103/PhysRevX.8.011021}
}

@article{chem-5,
	title = {Quantum Chemistry Calculations on a Trapped-Ion Quantum Simulator},
	author = {Hempel, Cornelius and Maier, Christine and Romero, Jonathan and McClean, Jarrod and Monz, Thomas and Shen, Heng and Jurcevic, Petar and Lanyon, Ben P. and Love, Peter and Babbush, Ryan and Aspuru-Guzik, Al\'an and Blatt, Rainer and Roos, Christian F.},
	journal = {Phys. Rev. X},
	volume = {8},
	issue = {3},
	pages = {031022},
	numpages = {22},
	year = {2018},
	month = {Jul},
	publisher = {American Physical Society},
	doi = {10.1103/PhysRevX.8.031022},
	url = {https://link.aps.org/doi/10.1103/PhysRevX.8.031022}
}

@misc{chem-6,
title={Ground-state energy estimation of the water molecule on a trapped ion quantum computer}, 
author={Yunseong Nam and Jwo-Sy Chen and Neal C. Pisenti and Kenneth Wright and Conor Delaney and Dmitri Maslov and Kenneth R. Brown and Stewart Allen and Jason M. Amini and Joel Apisdorf and Kristin M. Beck and Aleksey Blinov and Vandiver Chaplin and Mika Chmielewski and Coleman Collins and Shantanu Debnath and Andrew M. Ducore and Kai M. Hudek and Matthew Keesan and Sarah M. Kreikemeier and Jonathan Mizrahi and Phil Solomon and Mike Williams and Jaime David Wong-Campos and Christopher Monroe and Jungsang Kim},
year={2019},
eprint={1902.10171},
archivePrefix={arXiv},
url={https://arxiv.org/abs/1902.10171}, 
}

@article{chem-7,
	title = {Quantum computational chemistry},
	author = {McArdle, Sam and Endo, Suguru and Aspuru-Guzik, Al\'an and Benjamin, Simon C. and Yuan, Xiao},
	journal = {Rev. Mod. Phys.},
	volume = {92},
	issue = {1},
	pages = {015003},
	numpages = {51},
	year = {2020},
	month = {Mar},
	publisher = {American Physical Society},
	doi = {10.1103/RevModPhys.92.015003},
	url = {https://link.aps.org/doi/10.1103/RevModPhys.92.015003}
}

@Article{lattice-1,
	author={Kokail, C.
	and Maier, C.
	and van Bijnen, R.
	and Brydges, T.
	and Joshi, M. K.
	and Jurcevic, P.
	and Muschik, C. A.
	and Silvi, P.
	and Blatt, R.
	and Roos, C. F.
	and Zoller, P.},
	title={Self-verifying variational quantum simulation of lattice models},
	journal={Nature},
	year={2019},
	month={May},
	day={01},
	volume={569},
	number={7756},
	pages={355-360},
	issn={1476-4687},
	doi={10.1038/s41586-019-1177-4},
	url={https://doi.org/10.1038/s41586-019-1177-4}
}

@article{uccsd,
	doi = {10.1088/2058-9565/aad3e4},
	url = {https://dx.doi.org/10.1088/2058-9565/aad3e4},
	year = {2018},
	month = {oct},
	publisher = {IOP Publishing},
	volume = {4},
	number = {1},
	pages = {014008},
	author = {Jonathan Romero and Ryan Babbush and Jarrod R McClean and Cornelius Hempel and Peter J Love and Alán Aspuru-Guzik},
	title = {Strategies for quantum computing molecular energies using the unitary coupled cluster ansatz},
	journal = {Quantum Sci. Technol.}
}

@article{uccgsd,
	author = {Lee, Joonho and Huggins, William J. and Head-Gordon, Martin and Whaley, K. Birgitta},
	title = {Generalized Unitary Coupled Cluster Wave functions for Quantum Computation},
	journal = {J. Chem. Theory Comput.},
	volume = {15},
	number = {1},
	pages = {311-324},
	year = {2019},
	doi = {10.1021/acs.jctc.8b01004},
	
	URL = { 
	https://doi.org/10.1021/acs.jctc.8b01004
	
	}
	
}

@article{fixed-ansatz,
	title = {Progress towards practical quantum variational algorithms},
	author = {Wecker, Dave and Hastings, Matthew B. and Troyer, Matthias},
	journal = {Phys. Rev. A},
	volume = {92},
	issue = {4},
	pages = {042303},
	numpages = {10},
	year = {2015},
	month = {Oct},
	publisher = {American Physical Society},
	doi = {10.1103/PhysRevA.92.042303},
	url = {https://link.aps.org/doi/10.1103/PhysRevA.92.042303}
}

@article{symmetry,
	title = {Quantum algorithms for electronic structure calculations: Particle-hole Hamiltonian and optimized wave-function expansions},
	author = {Barkoutsos, Panagiotis Kl. and Gonthier, Jerome F. and Sokolov, Igor and Moll, Nikolaj and Salis, Gian and Fuhrer, Andreas and Ganzhorn, Marc and Egger, Daniel J. and Troyer, Matthias and Mezzacapo, Antonio and Filipp, Stefan and Tavernelli, Ivano},
	journal = {Phys. Rev. A},
	volume = {98},
	issue = {2},
	pages = {022322},
	numpages = {13},
	year = {2018},
	month = {Aug},
	publisher = {American Physical Society},
	doi = {10.1103/PhysRevA.98.022322},
	url = {https://link.aps.org/doi/10.1103/PhysRevA.98.022322}
}

@article{hva,
	title = {Exploring Entanglement and Optimization within the Hamiltonian Variational Ansatz},
	author = {Wiersema, Roeland and Zhou, Cunlu and de Sereville, Yvette and Carrasquilla, Juan Felipe and Kim, Yong Baek and Yuen, Henry},
	journal = {PRX Quantum},
	volume = {1},
	issue = {2},
	pages = {020319},
	numpages = {14},
	year = {2020},
	month = {Dec},
	publisher = {American Physical Society},
	doi = {10.1103/PRXQuantum.1.020319},
	url = {https://link.aps.org/doi/10.1103/PRXQuantum.1.020319}
}

@Article{adaptvqe,
	author={Grimsley, Harper R.
	and Economou, Sophia E.
	and Barnes, Edwin
	and Mayhall, Nicholas J.},
	title={An adaptive variational algorithm for exact molecular simulations on a quantum computer},
	journal={Nat. Commun.},
	year={2019},
	month={Jul},
	day={08},
	volume={10},
	number={1},
	pages={3007},
	issn={2041-1723},
	doi={10.1038/s41467-019-10988-2},
	url={https://doi.org/10.1038/s41467-019-10988-2}
}

@article{adaptvqe-nuc,
	title = {Solving nuclear structure problems with the adaptive variational quantum algorithm},
	author = {Romero, A. M. and Engel, J. and Tang, Ho Lun and Economou, Sophia E.},
	journal = {Phys. Rev. C},
	volume = {105},
	issue = {6},
	pages = {064317},
	numpages = {9},
	year = {2022},
	month = {Jun},
	publisher = {American Physical Society},
	doi = {10.1103/PhysRevC.105.064317},
	url = {https://link.aps.org/doi/10.1103/PhysRevC.105.064317}
}

@article{qubit-adaptvqe,
	title = {{Qubit-ADAPT-VQE}: An Adaptive Algorithm for Constructing Hardware-Efficient Ans\"atze on a Quantum Processor},
	author = {Tang, Ho Lun and Shkolnikov, V.O. and Barron, George S. and Grimsley, Harper R. and Mayhall, Nicholas J. and Barnes, Edwin and Economou, Sophia E.},
	journal = {PRX Quantum},
	volume = {2},
	issue = {2},
	pages = {020310},
	numpages = {16},
	year = {2021},
	month = {Apr},
	publisher = {American Physical Society},
	doi = {10.1103/PRXQuantum.2.020310},
	url = {https://link.aps.org/doi/10.1103/PRXQuantum.2.020310}
}

@article{iqcc,
	author = {Ryabinkin, Ilya G. and Lang, Robert A. and Genin, Scott N. and Izmaylov, Artur F.},
	title = {Iterative Qubit Coupled Cluster Approach with Efficient Screening of Generators},
	journal = {J. Chem. Theory Comput.},
	volume = {16},
	number = {2},
	pages = {1055-1063},
	year = {2020},
	doi = {10.1021/acs.jctc.9b01084},
	
	URL = { 
	https://doi.org/10.1021/acs.jctc.9b01084
	
	}
	
}

@article{cluster-vqe,
	author={Zhang, Yu
	and Cincio, Lukasz
	and Negre, Christian F. A.
	and Czarnik, Piotr
	and Coles, Patrick J.
	and Anisimov, Petr M.
	and Mniszewski, Susan M.
	and Tretiak, Sergei
	and Dub, Pavel A.},
	title={Variational quantum eigensolver with reduced circuit complexity},
	journal={npj Quantum Information},
	year={2022},
	month={Aug},
	day={12},
	volume={8},
	number={1},
	pages={96},
	issn={2056-6387},
	doi={10.1038/s41534-022-00599-z},
	url={https://doi.org/10.1038/s41534-022-00599-z}
}

@article{rotoselect,
	doi = {10.22331/q-2021-01-28-391},
	url = {https://doi.org/10.22331/q-2021-01-28-391},
	title = {Structure optimization for parameterized quantum circuits},
	author = {Ostaszewski, Mateusz and Grant, Edward and Benedetti, Marcello},
	journal = {{Quantum}},
	issn = {2521-327X},
	publisher = {{Verein zur F{\"{o}}rderung des Open Access Publizierens in den Quantenwissenschaften}},
	volume = {5},
	pages = {391},
	month = jan,
	year = {2021}
}

@article{vans,
author={Bilkis, M.
and Cerezo, M.
and Verdon, Guillaume
and Coles, Patrick J.
and Cincio, Lukasz},
title={A semi-agnostic ansatz with variable structure for variational quantum algorithms},
journal={Quantum Mach. Intell.},
year={2023},
month={Nov},
day={18},
volume={5},
number={2},
pages={43},
issn={2524-4914},
doi={10.1007/s42484-023-00132-1},
url={https://doi.org/10.1007/s42484-023-00132-1}
}

@misc{evo-vqe,
title={A Domain-agnostic, Noise-resistant, Hardware-efficient Evolutionary Variational Quantum Eigensolver}, 
author={Arthur G. Rattew and Shaohan Hu and Marco Pistoia and Richard Chen and Steve Wood},
year={2020},
eprint={1910.09694},
archivePrefix={arXiv},
 
url={https://arxiv.org/abs/1910.09694}, 
}

@misc{mog-vqe,
title={MoG-VQE: Multiobjective genetic variational quantum eigensolver}, 
author={D. Chivilikhin and A. Samarin and V. Ulyantsev and I. Iorsh and A. R. Oganov and O. Kyriienko},
year={2020},
eprint={2007.04424},
archivePrefix={arXiv},
 
url={https://arxiv.org/abs/2007.04424}, 
}

@article{qas,
author={Du, Yuxuan
and Huang, Tao
and You, Shan
and Hsieh, Min-Hsiu
and Tao, Dacheng},
title={Quantum circuit architecture search for variational quantum algorithms},
journal={npj Quantum Inf.},
year={2022},
month={May},
day={23},
volume={8},
number={1},
pages={62},
issn={2056-6387},
doi={10.1038/s41534-022-00570-y},
url={https://doi.org/10.1038/s41534-022-00570-y}
}

@misc{morse,
title={Morse Potential on a Quantum Computer for Molecules and Supersymmetric Quantum Mechanics}, 
author={Josh Apanavicius and Yuan Feng and Yasmin Flores and Mohammad Hassan and Michael McGuigan},
year={2021},
eprint={2102.05102},
archivePrefix={arXiv},
 
url={https://arxiv.org/abs/2102.05102}, 
}

@Article{dvr-dynamics,
	author={Lee, Chee-Kong
	and Hsieh, Chang-Yu
	and Zhang, Shengyu
	and Shi, Liang},
	title={Variational Quantum Simulation of Chemical Dynamics with Quantum Computers},
	journal={J. Chem. Theory Comput.},
	year={2022},
	month={Apr},
	day={12},
	publisher={American Chemical Society},
	volume={18},
	number={4},
	pages={2105-2113},
	issn={1549-9618},
	doi={10.1021/acs.jctc.1c01176},
	url={https://doi.org/10.1021/acs.jctc.1c01176}
}

@Article{mol-vib,
	author ="McArdle, Sam and Mayorov, Alexander and Shan, Xiao and Benjamin, Simon and Yuan, Xiao",
	title  ="Digital quantum simulation of molecular vibrations",
	journal  ="Chem. Sci.",
	year  ="2019",
	volume  ="10",
	issue  ="22",
	pages  ="5725-5735",
	publisher  ="The Royal Society of Chemistry",
	doi  ="10.1039/C9SC01313J",
	url  ="http://dx.doi.org/10.1039/C9SC01313J"}

@Article{mol-vib-2,
	author ="Ollitrault, Pauline J. and Baiardi, Alberto and Reiher, Markus and Tavernelli, Ivano",
	title  ="Hardware efficient quantum algorithms for vibrational structure calculations",
	journal  ="Chem. Sci.",
	year  ="2020",
	volume  ="11",
	issue  ="26",
	pages  ="6842-6855",
	publisher  ="The Royal Society of Chemistry",
	doi  ="10.1039/D0SC01908A",
	url  ="http://dx.doi.org/10.1039/D0SC01908A"}

@article{mol-vib-3,
	title = {Near- and long-term quantum algorithmic approaches for vibrational spectroscopy},
	author = {Sawaya, Nicolas P. D. and Paesani, Francesco and Tabor, Daniel P.},
	journal = {Phys. Rev. A},
	volume = {104},
	issue = {6},
	pages = {062419},
	numpages = {15},
	year = {2021},
	month = {Dec},
	publisher = {American Physical Society},
	doi = {10.1103/PhysRevA.104.062419},
	url = {https://link.aps.org/doi/10.1103/PhysRevA.104.062419}
}

@article{cr2-pot,
	title = {Magnetic Feshbach resonances and Zeeman relaxation in bosonic chromium gas with anisotropic interaction},
	author = {Pavlovi\ifmmode \acute{c}\else \'{c}\fi{}, Z. and Krems, R. V. and C\^ot\'e, R. and Sadeghpour, H. R.},
	journal = {Phys. Rev. A},
	volume = {71},
	issue = {6},
	pages = {061402},
	numpages = {4},
	year = {2005},
	month = {Jun},
	publisher = {American Physical Society},
	doi = {10.1103/PhysRevA.71.061402},
	url = {https://link.aps.org/doi/10.1103/PhysRevA.71.061402}
}

@article{arhcl-pot,
	author = {Hutson, Jeremy M.},
	title = {Vibrational dependence of the anisotropic intermolecular potential of argon-hydrogen chloride},
	journal = {J. Phys. Chem.},
	volume = {96},
	number = {11},
	pages = {4237-4247},
	year = {1992},
	doi = {10.1021/j100190a026},
	
	URL = { 
	https://doi.org/10.1021/j100190a026
	
	}
	
}

@Article{mgnh-pot,
	author ="Soldán, Pavel and Żuchowski, Piotr S. and Hutson, Jeremy M.",
	title  ="Prospects for sympathetic cooling of polar molecules: {NH} with alkali-metal and alkaline-earth atoms – a new hope",
	journal  ="Faraday Discuss.",
	year  ="2009",
	volume  ="142",
	issue  ="0",
	pages  ="191-201",
	publisher  ="The Royal Society of Chemistry",
	doi  ="10.1039/B822769C",
	url  ="http://dx.doi.org/10.1039/B822769C"}

@article{vqd,
	doi = {10.22331/q-2019-07-01-156},
	url = {https://doi.org/10.22331/q-2019-07-01-156},
	title = {Variational {Q}uantum {C}omputation of {E}xcited {S}tates},
	author = {Higgott, Oscar and Wang, Daochen and Brierley, Stephen},
	journal = {{Quantum}},
	issn = {2521-327X},
	publisher = {{Verein zur F{\"{o}}rderung des Open Access Publizierens in den Quantenwissenschaften}},
	volume = {3},
	pages = {156},
	month = jul,
	year = {2019}
}

@misc{vqe-review,
	doi = {10.48550/ARXIV.2111.05176},
	
	archiveprefix = {arXiv},
	eprint = {2111.05176},
	url = {https://arxiv.org/abs/2111.05176},
	
	author = {Tilly, Jules and Chen, Hongxiang and Cao, Shuxiang and Picozzi, Dario and Setia, Kanav and Li, Ying and Grant, Edward and Wossnig, Leonard and Rungger, Ivan and Booth, George H. and Tennyson, Jonathan},
	
	title = {The Variational Quantum Eigensolver: a review of methods and best practices},
	
	publisher = {arXiv},
	
	year = {2021},
	
	copyright = {Creative Commons Attribution 4.0 International}
}

@article{choi,
	author = {Choi,Seung E.  and Light,J. C. },
	title = {Determination of the bound and quasibound states of {Ar–HCl} van der Waals complex: Discrete variable representation method},
	journal = {J. Chem. Phys.},
	volume = {92},
	number = {4},
	pages = {2129-2145},
	year = {1990},
	doi = {10.1063/1.458004},
	
	URL = { 
	https://doi.org/10.1063/1.458004
	
	}
	
}

@article{colbert,
	author = {Colbert,Daniel T.  and Miller,William H. },
	title = {A novel discrete variable representation for quantum mechanical reactive scattering via the S‐matrix Kohn method},
	journal = {J. Chem. Phys.},
	volume = {96},
	number = {3},
	pages = {1982-1991},
	year = {1992},
	doi = {10.1063/1.462100},
	
	URL = { 
	https://doi.org/10.1063/1.462100
	
	}	
}

@article{arhcl-obs-0,
	author = {Pine,A. S.  and Howard,B. J. },
	title = {Hydrogen bond energies of the {HF} and {HCl} dimers from absolute infrared intensities},
	journal = {J. Chem. Phys.},
	volume = {84},
	number = {2},
	pages = {590-596},
	year = {1986},
	doi = {10.1063/1.450605},
	
	URL = { 
	https://doi.org/10.1063/1.450605
	
	}
	
}

@article{arhcl-obs-1,
	title = {Tunable far-infrared laser spectroscopy in a planar supersonic jet: The $\Sigma$ bending vibration of {Ar${}^{35}$HCl}},
	journal = {Chem. Phys. Lett.},
	volume = {141},
	number = {4},
	pages = {289-291},
	year = {1987},
	issn = {0009-2614},
	doi = {https://doi.org/10.1016/0009-2614(87)85025-X},
	url = {https://www.sciencedirect.com/science/article/pii/000926148785025X},
	author = {Kerry L. Busarow and Geoffrey A. Blake and K.B. Laughlin and R.C. Cohen and Y.T. Lee and R.J. Saykally},
}

@article{arhcl-obs-12,
	doi       = {10.1080/00268978800100741},
	title     = {Far infrared laser Stark spectroscopy of the $\Sigma$ bending vibration of {ArHCl}},
	author    = {Robinson, Ruth L. and Gwo, Dz-Hung and Saykally, Richard J.},
	publisher = {Taylor and Francis Group},
	journal   = {Mol. Phys.},
	issn      = {0026-8976,1362-3028},
	year      = {1988},
	volume    = {63},
	issue     = {6},
	pages     = {1021--1029},
	url       = {http://doi.org/10.1080/00268978800100741}
}

@article{arhcl-obs-2,
	author = {Robinson,Ruth L.  and Gwo,Dz‐Hung  and Saykally,Richard J. },
	title = {The high‐resolution far infrared spectrum of a van der Waals stretching vibration: The $\nu_3$ band of Ar–HCl},
	journal = {J. Chem. Phys.},
	volume = {87},
	number = {9},
	pages = {5156-5160},
	year = {1987},
	doi = {10.1063/1.453684},
	
	URL = { 
	https://doi.org/10.1063/1.453684
	
	}
	
}

@article{co2-1,
	author = {Lötstedt,Erik  and Yamanouchi,Kaoru  and Tachikawa,Yutaka },
	title = {Evaluation of vibrational energies and wave functions of {CO}${}_2$ on a quantum computer},
	journal = {AVS Quantum Sci.},
	volume = {4},
	number = {3},
	pages = {036801},
	year = {2022},
	doi = {10.1116/5.0091144},
	
	URL = { 
	https://doi.org/10.1116/5.0091144
	
	}
	
}

@article{co2-2,
	title = {Calculation of vibrational eigenenergies on a quantum computer: Application to the Fermi resonance in {CO}${}_2$},
	author = {L\"otstedt, Erik and Yamanouchi, Kaoru and Tsuchiya, Takashi and Tachikawa, Yutaka},
	journal = {Phys. Rev. A},
	volume = {103},
	issue = {6},
	pages = {062609},
	numpages = {12},
	year = {2021},
	month = {Jun},
	publisher = {American Physical Society},
	doi = {10.1103/PhysRevA.103.062609},
	url = {https://link.aps.org/doi/10.1103/PhysRevA.103.062609}
}

@article{co2-3,
	author = {Nguyen, Manh Tien and Lee, Yueh-Lin and Alfonso, Dominic and Shao, Qing and Duan, Yuhua},
	title = "{Description of reaction and vibrational energetics of CO2–NH3 interaction using quantum computing algorithms}",
	journal = {AVS Quantum Science},
	volume = {5},
	number = {1},
	pages = {013801},
	year = {2023},
	month = {03},
	issn = {2639-0213},
	doi = {10.1116/5.0137750},
	url = {https://doi.org/10.1116/5.0137750},
}

@Article{comp-search-1,
	author={Grimsley, Harper R.
	and Economou, Sophia E.
	and Barnes, Edwin
	and Mayhall, Nicholas J.},
	title={An adaptive variational algorithm for exact molecular simulations on a quantum computer},
	journal={Nat. Commun.},
	year={2019},
	month={Jul},
	day={08},
	volume={10},
	number={1},
	pages={3007},
	issn={2041-1723},
	doi={10.1038/s41467-019-10988-2},
	url={https://doi.org/10.1038/s41467-019-10988-2}
}

@article{comp-search-2,
	title = {Compositional optimization of quantum circuits for quantum kernels of support vector machines},
	author = {Torabian, Elham and Krems, Roman V.},
	journal = {Phys. Rev. Res.},
	volume = {5},
	issue = {1},
	pages = {013211},
	numpages = {8},
	year = {2023},
	month = {Mar},
	publisher = {American Physical Society},
	doi = {10.1103/PhysRevResearch.5.013211},
	url = {https://link.aps.org/doi/10.1103/PhysRevResearch.5.013211}
}

@article{demler,
	title = {Ab initio exact diagonalization simulation of the Nagaoka transition in quantum dots},
	author = {Wang, Yao and Dehollain, Juan Pablo and Liu, Fang and Mukhopadhyay, Uditendu and Rudner, Mark S. and Vandersypen, Lieven M. K. and Demler, Eugene},
	journal = {Phys. Rev. B},
	volume = {100},
	issue = {15},
	pages = {155133},
	numpages = {14},
	year = {2019},
	month = {Oct},
	publisher = {American Physical Society},
	doi = {10.1103/PhysRevB.100.155133},
	url = {https://link.aps.org/doi/10.1103/PhysRevB.100.155133}
}

@article{xuyang,
	doi = {10.1088/2632-2153/ad7cc1},
	url = {https://dx.doi.org/10.1088/2632-2153/ad7cc1},
	year = {2024},
	month = {sep},
	publisher = {IOP Publishing},
	volume = {5},
	number = {3},
	pages = {035081},
	author = {Xuyang Guo and Jun Dai and Roman V Krems},
	title = {Benchmarking of quantum fidelity kernels for Gaussian process regression},
	journal = {Mach. Learn.: Sci. Technol.}
}

@misc{sm,
	title = {Supplementary materials}
}
	
\end{document}